\documentclass[12pt,indent]{article}%a5paper
\usepackage{a4,latexsym,amsmath,amssymb}%a4
\usepackage[latin1]{inputenc}
\usepackage{float}
\usepackage{natbib}
\usepackage{graphics}
\usepackage[english]{babel}
\usepackage{tabularx}
\usepackage{lscape}
\usepackage{multirow}
\usepackage{rotating}
\usepackage{dsfont}
\usepackage{subfig}
\usepackage{bbm}
\usepackage{arydshln}
\usepackage{booktabs}
\usepackage[table]{xcolor}

\usepackage{calc}
\usepackage{ifthen}
\newenvironment{tabularsmall}
{ \footnotesize \sffamily \tabular } {
\endtabular
\normalfont }
\newcommand{\betab}{{\boldsymbol{\beta}}}

\newcommand{\phib}{{\boldsymbol{\phi}}}

\newcommand{\Sigmab}{\boldsymbol{\mathit\Sigma}}

\newcommand{\bb}{\boldsymbol{b}}

\newcommand{\xb}{\boldsymbol{x}}

\newcommand{\zb}{\boldsymbol{z}}

\newcommand{\0}{\textbf{0}}
\newcommand{\blanco}[1]{}

\def\d{\displaystyle}

\usepackage{natbib}

\usepackage{setspace}
\begin{document}
\bibliographystyle{chicago}
\sloppy
%%%%%%%%%%%%%%%%%%%%%%%%%%%%%%%%%%%%%%%%%%%%%%%%%%%%%%%%%%%%%%%%%%%%%
%                                                                   %
%         Definition einer modifizierten Kapitelï¿½berschrift         %
%                                                                   %

\makeatletter
\renewcommand{\section}{\@startsection{section}{1}{\z@}%
        {-3.5ex \@plus -1ex \@minus -.2ex}%
        {1.5ex \@plus.2ex}%
        {\reset@font\Large\sffamily}}
\renewcommand{\subsection}{\@startsection{subsection}{1}{\z@}%
        {-3.25ex \@plus -1ex \@minus -.2ex}%
        {1.1ex \@plus.2ex}%
        {\reset@font\large\sffamily\flushleft}}
\renewcommand{\subsubsection}{\@startsection{subsubsection}{1}{\z@}%
        {-3.25ex \@plus -1ex \@minus -.2ex}%
        {1.1ex \@plus.2ex}%
        {\reset@font\normalsize\sffamily\flushleft}}
\makeatother

%                                                                   %
%%%%%%%%%%%%%%%%%%%%%%%%%%%%%%%%%%%%%%%%%%%%%%%%%%%%%%%%%%%%%%%%%%%%%

%%%%%%%%%%%%%%%%%%%%%%%%%%%%%%%%%%%%%%%%%%%%%%%%%%%%%%%%%%%%%%%%%%%%%
%                                                                   %
%         Definition einer modifizierten Bildunterschrift           %
%                                                                   %

\newsavebox{\tempbox}
\newlength{\linelength}
\setlength{\linelength}{\linewidth-10mm} \makeatletter
\renewcommand{\@makecaption}[2]
{
  \renewcommand{\baselinestretch}{1.1} \normalsize\small
  \vspace{5mm}
  \sbox{\tempbox}{#1: #2}
  \ifthenelse{\lengthtest{\wd\tempbox>\linelength}}
  {\noindent\hspace*{4mm}\parbox{\linewidth-10mm}{\sc#1: \sl#2\par}}
  {\begin{center}\sc#1: \sl#2\par\end{center}}
}

%                                                                   %
%%%%%%%%%%%%%%%%%%%%%%%%%%%%%%%%%%%%%%%%%%%%%%%%%%%%%%%%%%%%%%%%%%%%%

%\bibliographystyle{chicago}
%\baselineskip7mm
%\parindent 0.5cm
%\parskip2ex plus0.5ex minus 0.5ex
%\setlength{\parskip}{7pt plus 1pt minus 1pt}

\def\R{\mathchoice{ \hbox{${\rm I}\!{\rm R}$} }
                   { \hbox{${\rm I}\!{\rm R}$} }
                   { \hbox{$ \scriptstyle  {\rm I}\!{\rm R}$} }
                   { \hbox{$ \scriptscriptstyle  {\rm I}\!{\rm R}$} }  }

\def\N{\mathchoice{ \hbox{${\rm I}\!{\rm N}$} }
                   { \hbox{${\rm I}\!{\rm N}$} }
                   { \hbox{$ \scriptstyle  {\rm I}\!{\rm N}$} }
                   { \hbox{$ \scriptscriptstyle  {\rm I}\!{\rm N}$} }  }

\def\d{\displaystyle}

\title{Tree-Structured Clustering in Fixed Effects Models}
\author{Moritz Berger \& Gerhard Tutz \\{\small Ludwig-Maximilians-Universit\"{a}t M\"{u}nchen}\\
{\small Akademiestra{\ss}e 1, 80799 M\"{u}nchen}}

%\author{Jan Gertheiss\footnote{To whom correspondence should be
%addressed: \texttt{jan.gertheiss@stat.uni-muenchen.de.}}
%\footnote{Department of Statistics, Ludwig-Maximilians-Universität
%Munich, Germany.} \ \& Gerhard Tutz\footnotemark[2]}

% \ead{tutz@stat.uni-muenchen.de}
% \address{Ludwig-Maximilians-Universit\"{a}t M\"{u}nchen, Ludwigstra{\ss}e 33, D-80539 M\"{u}nchen, Germany}
%\author{Lorenz Uhlmann}
%\ead{tutz@stat.uni-muenchen.de}
%\address[muc]{Ludwig-Maximilians-Universit\"{a}t M\"{u}nchen, Akademiestra{\ss}e 1, 80799 M\"{u}nchen, Germany}
%\cortext[cor]{Corresponding author. Tel.: ++4989 2180 3044; fax.:
%++4989 2180 5308.}
%{\texttt{\small \{tutz, uhlmann\}@stat.uni-muenchen.de}}}
%\address[muc1]{Ludwig-Maximilians-University Munich, Ludwigstrasse 33, D-80539 Munich, Germany}
%\address[muc2]{Ludwig-Maximilians-University Munich, Akademiestra{\ss}e 1, D-80799 Munich, Germany}
%\cortext[cor]{Corresponding author. Tel.: ++49 89 2180 3044; fax.:
%++49 89 2180 ???.}
\maketitle

\begin{abstract}
\noindent
Fixed effects models are very flexible because they do not make assumptions on the distribution of effects and can also be used if the heterogeneity component is correlated with explanatory variables. A disadvantage is the large number of effects that have to be estimated. A recursive partitioning (or tree based) method is proposed that identifies clusters of units that share the same effect. The approach reduces the number of parameters to be estimated and is useful in particular if one is interested in identifying clusters with the same  effect on a response variable. It is shown that the method performs well and outperforms competitors like the finite mixture model  in particular if the heterogeneity component is correlated with explanatory variables. In two applications  the usefulness of the approach to identify clusters that share the same effect is illustrated.

\end{abstract}


\section{Introduction}\label{Introduction}

\blanco{
Clustered data are found in cross-sectional studies with multistage sampling of measurement units. As an example we consider data of a multi-center treatment study that investigates the effect of beta blockers after myocardial infarction. Clustered data are also obtained in longitudinal studies, where each unit is measured repeatedly over time. In a second example we consider data of the German Socio-Economic Panel (SOEP), which contain annual observations of the participants for the period from 2000 to 2012. Following \citet{WunWien2013} our main attention is to investigate the satisfaction of the participants over the life span.
}

The analysis of longitudinal data and cross-sectional data that come in clusters requires to take the dependence of observations and the heterogeneity of measurement units into account. Typically, measurements within units tend to be more similar than measurements between units. If the heterogeneity is ignored poor performance of estimators and misleading standard errors are to be expected.

The most popular, widely used model to account for unobserved heterogeneity is the random effects model, see, for example, \citet{VerMol:2000}, \citet{MolVer:2005} and \citet{MccSea:2001}. Typically in the random effects model it is assumed that the random effects follow a normal distribution. This strong assumption results in an economical model but  inference may be sensitive to the specification of the distribution of random effects, see \citet{HeaKur:2001}, \citet{AgrCafOhm:2004} and \citet{LitAloMol:2007}.

%An alternative approach, which is considered here, are fixed effects models. In fixed effects models one single parameter for each measurement unit is specified. An advantage is that  no structural assumptions on the unit-specific effects have to be made, a drawback is that the number of parameters to be estimated is inflated.

Several approaches to weaken the assumption of normally distributed random effects have been proposed.
More flexible distributions are obtained, for example, by using mixtures of normals as proposed by \citet{Chenetal:2002} and
\citet{MagZeg:96}. \citet{Huang:2009}  proposed diagnostic methods for
random-effect misspecification and \citet{ClaHart:2009}  proposed tests for the assumption of the normal distribution. More recently, \citet{lombardia2012new} proposed the class of semi-mixed effects models, a continuum of models that combine random and fixed effects. 

An alternative approach to model heterogeneity uses finite mixtures. In finite mixtures of generalized linear models it is assumed that the density or mass function of  the responses given the explanatory variables is determined by a finite mixture of components. Each of the components has its own response distribution and own parameters that determine the influence of explanatory variables. If only part of the parameters, for example the intercepts, are allowed to vary over components one obtains a discrete distribution of the heterogeneity part of the model.
Models of that type were considered by \citet{FolLam:89} and \citet{Aitkin:99}.
\citet{FolLam:89} investigated the identifiability of finite mixtures of binomial regression models and gave sufficient
identifiability conditions for mixing of binary and  binomial distributions. \citet{grun2008identifiability} considered identifiability for mixtures of multinomial logit models.

Finite mixture models replace the assumption of a fixed continuous distribution of random effects by the assumption of a discrete distribution. One may see this as an alternative and flexible specification of the heterogeneity component only. However, by assuming a discrete distribution of the intercepts instead of a continuous distribution as in random effects models one also implicitly assumes that there are clusters of units that share the same effect. In some applications it is definitely of interest to identify these units. We will consider an example in which the units are schools and one wants to know which schools are similar in their performance with regard to the education of students.

Here we consider an alternative to finite mixture models with the same objectives, that are use of a flexible discrete distribution and  identification of units that share the same effect.
However, the starting point is different. We use a fixed effects model in which each unit has its own parameter. An advantage is that  no structural assumptions on the unit-specific effects have to be made. Clusters of parameters and therefore units with the same effect are found by tree methodology, although a different one as in classical trees.

Classical recursive partitioning techniques or trees were first introduced by \citet{MorSon:63}. Very popular methods are classification and regression trees (CART) by \citet{BreiFrieOls:84} and C4.5 by \citet{Quinlan:86} and \citet{Quinlan:93}. A newer version of recursive partitioning based on conditional inference was proposed by \citet{Hotetal:2006}. An overview on recursive partitioning in health science was given by \citet{ZhaSin:1999} and with a focus on psychometrics by \citet{Strobetal:2009}. An easily accessible  introduction into the basic concepts is found in  \citet{HasTibFri:2009B}.

The tree methodology used here differs from these approaches. In CART and other classical approaches the whole covariate space is recursively partitioned into subspaces. In order to obtain a partitioning in the intercepts (or slopes) only, one has to apply a different form of trees. It has to be designed in a way that the  subspaces are built for specific effects only, for example the intercepts,  while other parameters that represent common effects of explanatory variables are not partitioned into subspaces. Our main focus is on the clustering of intercepts, however, we will also refer to the case of unit-specific slopes. One big advantage using recursive partitioning techniques is the computational efficiency. The proposed tree-structured model especially enables the evaluation of high-dimensional data. Alternative approaches to identify clusters within a fixed effects model framework as proposed by \citet{TuOelkFixed} fail in high dimensional settings.

The article is organized as follows: In Section \ref{sec:model} we introduce the tree-structured model for unit-specific intercepts and in section \ref{sec:example} we present an illustrative example. Details about the fitting procedure are given in Section \ref{sec:fitting procedure}. After a short introduction of related approaches in Section \ref{sec:related} we give the results of wider simulation studies (Section \ref{sec:simulations}). Finally, Section \ref{sec:application} contains a second application. 
%Finally, in Section \ref{sec:extension} we consider the extension to models with unit-specific slopes.

\section{Accounting for Heterogeneity in Clustered Data} \label{sec:model}

Consider clustered data given by $(y_{ij},\xb_{ij},\zb_{ij}),\,i=1,\hdots,n,\,j=1,\hdots,n_i$, where $y_{ij}$ denotes the response of measurement $j$ for unit $i$ and two sets of predictive variables $\xb_{ij}^\top=(1,x_{ij1},\hdots,x_{ijp})$ and $\zb_{ij}^\top=(1,z_{ij1},\hdots,z_{ijq})$. In longitudinal data the units can, for example, represent persons that are measured repeatedly. In the following, we consider alternative methods to account for the potential heterogeneity of units. We start with methods
that use random effects, then consider fixed effects model and finite mixtures.

\subsection{Random Effects Models}

In a generalized linear mixed model (GLMM) the mean response $\mu_{ij}=\mathbb{E}(y_{ij}|\bb_i,\xb_{ij},\zb_{ij})$ is linked to the explanatory variables by
\begin{equation}
g(\mu_{ij})=\xb_{ij}^\top\betab+\zb_{ij}^\top\bb_i,
\end{equation}
where $\xb_{ij}^\top\betab$ is a linear term which contains the fixed effect  $\betab$. The second term $\zb_{ij}^\top\bb_i$ contains the random effects for covariates $\zb_{ij}$ that are varying across units and $g(\cdot)$ is a known link function. In a GLMM it is assumed that the distribution of $y_{ij}|\bb_i,\xb_{ij},\zb_{ij}$ follows a simple exponential family and that the observations $y_{ij}$ are conditionally independent. For the random effects $\bb_i$, which model the heterogeneity of the units, one typically assumes a normal distribution $\bb_i\sim N(\0,\Sigmab_{rand})$.

In a GLMM the distribution of the random effects is used to account for the heterogeneity of the units and the focus is mainly on the parametric term $\xb_{ij}^\top\betab$. Although the distributional assumption for the random effects makes the estimation of the model very efficient there are also some disadvantages. If the assumed distribution is very different from the real data generating distribution, inference can be biased. The assumption of a continuous distribution also does not allow for the same effects of different units. Hence, clustering of units is not possible. Another crucial point of the GLMM is the assumption that the random effects $\bb_i$ and the covariates $\xb_{ij}$ are uncorrelated. This assumption can lead to poor estimation accuracy, see, for example, \citet{grilli2011endo}. Functions for the estimation of generalized linear mixed models are provided by the R-package \texttt{lme4} \citep{lme4:2015}, which we will use for the computations in the applications and simulations.

\subsection{Fixed Effects Models}

In contrast to mixed models, fixed effects models model heterogeneity among units by using one parameter $\betab_i$ for each unit. The mean response $\mu_{ij}=\mathbb{E}(y_{ij}|\xb_{ij},\zb_{ij})$ is linked to the explanatory variables in the form
\begin{equation} \label{eq:fixedeffects_full}
g(\mu_{ij})=\eta_{ij}=\xb_{ij}^\top\betab+\zb_{ij}^\top\betab_i,
\end{equation}
where $\xb_{ij}$ again is a vector of covariates that have the same effect across all units and $\zb_{ij}$ contains covariates that have different effects over units. Each measurement unit has his own parameter vector $\betab_i^\top=(\beta_{i0},\hdots,\beta_{iq})$. The specification of one parameter vector per unit results in a very large number of parameters which can affect estimation accuracy. Moreover, typically there is  not enough information to distinguish between all units. To cope with these problems one can assume that there are groups of units that share the same effect on the response. Forming clusters of units leads to a reduced number of parameters and stable estimates. There are several strategies to identify these clusters, the fixed effects model with regularization considered in the next section or the finite mixture model (Section \ref{subsec:finite}).

\subsection{Tree-Structured Clustering} \label{subsec:tsc}

In the approach considered here one assumes that the fixed effects model holds, but not all the unit-specific parameters are assumed to be different. Clusters (or groups) of measurement
units are identified by recursive partitioning methods. We first consider unit-specific intercepts only. Let us start with the simplest case in which all intercepts are equal, that is, the linear predictor has the form $\eta_{ij}=\xb_{ij}^\top\betab+\beta_{0}$. If there are two clusters the corresponding linear predictor   is  given by
\begin{equation} \label{eq:fixedeffects_two}
\eta_{ij}=\xb_{ij}^\top\betab+\beta_{i0}^{(k)},\quad k=1,2,
\end{equation}
where $k$ denotes if the unit is in the first or the second group. A simple test, for example a likelihood ratio test, for the hypothesis $H_0:$ $\beta_{i0}^{(1)}=\beta_{i0}^{(2)}$ can be used to determine if the model with two groups is more adequate for the data than the model in which all the intercepts are equal.
%On the other hand, if more than two groups of units, let's say m groups, have to be distinguished, one has to specify m intercepts $\beta_{0i}^{(k)},\;k=1,\hdots,m$. The fitting procedure considered in the following is based on this simple model.
By iterative splitting into subsets guided by test statistics one obtains a clustering of units that have to be distinguished with regard to their intercept.

In general, regression trees can be seen as a representation of a partition of the predictor space. A tree is built by successively splitting one node $A$, that is already a subset of the predictor space, into two subsets $A_1$ and $A_2$ with the split being  determined by only one variable.
In a fixed effects model, when specifying specific intercepts for each unit, the unit number itself can be seen as a nominal categorical variable with $n$ categories. The partition has the form $A \cap S_1, \quad A \cap S_2$, where $S_1$ and $S_2$ are disjoint, non-empty subsets $S_1\subset\{1,\ldots,n\}$ and its complement $S_2=\{1,\ldots,n\}\setminus S_1$. Using this notation another representation of model \eqref{eq:fixedeffects_two} is given by
\begin{equation*} \label{eq:fixedeffects_two2}
\eta_{ij}=\xb_{ij}^\top\betab+\beta_{i0}^{(1)}I(i\in S_{10})+\beta_{i0}^{(2)}I(i\in S_{20}),
\end{equation*}
where $I(\cdot)$ denotes the indicator function with $I(a)=1$, if a is true and $I(a)=0$ otherwise.
After several splits one obtains a clustering of the units $\{1,\ldots,n\}$ and the predictor of the resulting model can be represented by
\begin{equation} \label{eq:fixedeffects_m0}
\eta_{ij}=\xb_{ij}^\top\betab+\sum_{k=1}^{m_0}{\beta_{i0}^{(k)}I(i\in S_{k0})}%=\xb_{ij}^\top\betab+\text{tr}(z_{ij0}),
\end{equation}
%\begin{align*}
%\eta_{ij}&=\xb_{ij}^\top\betab+\beta_{0i}^{(1)}I(i\in S_{10})+\hdots+\beta_{0i}^{(m)}I(i\in S_{m0})\\
%&=\xb_{ij}^\top\betab+tr(1)=\\
%&=\xb_{ij}^\top\betab+\sum_{\ell=1}^{m}{\beta_{0i}^{(\ell)}I(i\in S_{1\ell})}
%\end{align*}
where $S_{10},\hdots,S_{m_00}$ is a partition of $\{1,\hdots,n\}$ consisting of $m_0$ clusters that have to be distinguished in terms of their individual intercepts.

In the following we will use the model abbreviation \textit{TSC} for tree-structured clustering. 

%In the general case that there are additional variables whose effects differ among units the tree component for the r-th variable is given by
%\begin{equation}
%\text{tr}(z_{ijr})=\sum_{\ell=1}^{m_r}{z_{ijr}\beta_{ir}^{(\ell)}I(i\in S_{\ell r})}
%\end{equation}
%and the whole predictor of the resulting model can be written as
%\begin{equation}
%\eta_{ij}=\xb_{ij}^\top\betab+\text{tr}(z_{ij0})+\hdots+\text{tr}(z_{ijq})=\xb_{ij}^\top\betab+\text{tr}(\zb_{ij}).
%\end{equation}
%It is important that the number and form of partitions $S_{1r},\hdots,S_{m_rr},r=1,\hdots,q$ of $\{i,\hdots,n\}$ has not to be the same for each component of $\zb$.

\subsection{Finite Mixture Models} \label{subsec:finite}

An alternative approach that also allows to identify clusters of units are finite mixture models. These were, for example, considered by \citet{FolLam:89} and \citet{Aitkin:99}. The general assumption in finite mixtures of generalized regression models is that the mixture consists of K components where each component follows a parametric distribution of the exponential family of distributions. The density of the mixture can be given by
\begin{equation*}
f(y|\xb,\betab,\phib)=\sum_{k=1}^{K}{\pi_kf_k(y|\xb,\betab_k,\phi_k)}
\end{equation*}
where $f_k(y|\xb,\betab_k,\phi_k)$ denotes the $k$-th component of the mixture with parameter vector $\betab_k$ and dispersion parameter $\phi_k$. For the unknown component weights $\pi_k$
%can depend on an additional vector of covariates $\wb$, so called concomitant variables, and it holds
$\sum_{k=1}^{K}{\pi_k}=1$ and $\pi_k>0,\,k=1,\hdots,K$ has to hold.

Here we consider models with components that differ in  their intercepts. Within the framework of finite mixtures one specifies for the $k-$th component of the mixture  a model with predictor $\eta_{ij}^{(k)}=\beta_{i0}^{(k)}+x_{ij}^\top\betab$. For models with normal response the mixture components are given by $N(y_{ij}|\eta_{ij}^{(k)},\sigma^2)$, where the variance $\sigma^2$ is fixed for all components. For models with a binary response the mixture components are $B(y_{ij}|n,\pi_{ij}^{(k)})$, where $\pi_{ij}^{(k)} \in (0,1)$ and logit($\pi_{ij}^{(k)})=\eta_{ij}^{(k)}$. For further details, see \citet{GruLei:2007}.

Estimation of the mixture model is usually obtained  by the EM-algorithm with the number of components $K$ being specified beforehand. The optimal number of components is chosen afterwards, for example by information criteria like AIC or BIC.
\citet{GruLei:2008} provide  the R-package \texttt{flexmix}, which is used for the computations in our applications and simulations.
Regularization and variable selection for mixture models have been  considered by \citet{khalili07} and \citet{stadler10} but not with the objective of clustering units with regard to their effects.

\begin{table}[!ht]
\caption{Summary statistics of the test score of the 56 multiple-choice items and covariate gender of the CTB data of the illustrative example (CTB data).}
\begin{center}
\begin{tabularsmall}{lcccccc}
\toprule
Variable&\multicolumn{6}{c}{Summary statistics}\\
\midrule
&$x_{min}$&$x_{0.25}$&$x_{med}$&$\bar{x}$&$x_{0.75}$&$x_{max}$\\
Test score&21&32&34&34.14&37&46\\
\\
Gender&\multicolumn{3}{c}{male: 761}&\multicolumn{3}{c}{female: 739}\\
\bottomrule
\end{tabularsmall}
\end{center}
\label{tab:app_ctb_summary}
\end{table}

\begin{figure}[!ht]
\centering
\includegraphics[width=0.6\textwidth]{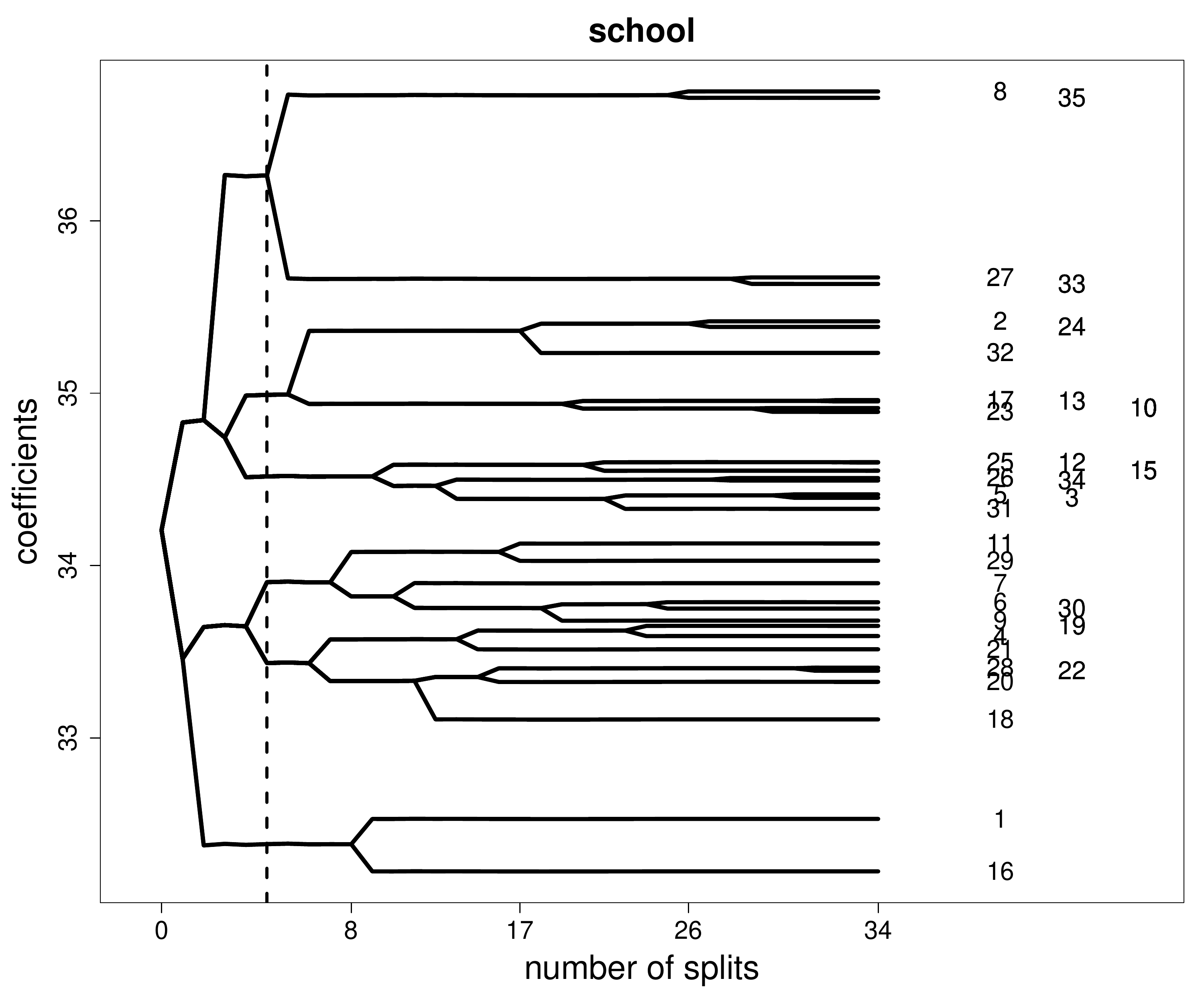}
\caption{Paths of coefficients of school-specific intercepts against all splits of the illustrative example (CTB data). The optimal number of splits is marked by a dashed line.}
\label{fig:results_ctb}
\end{figure}

\begin{figure}[!ht]
\centering
\includegraphics[width=0.7\textwidth]{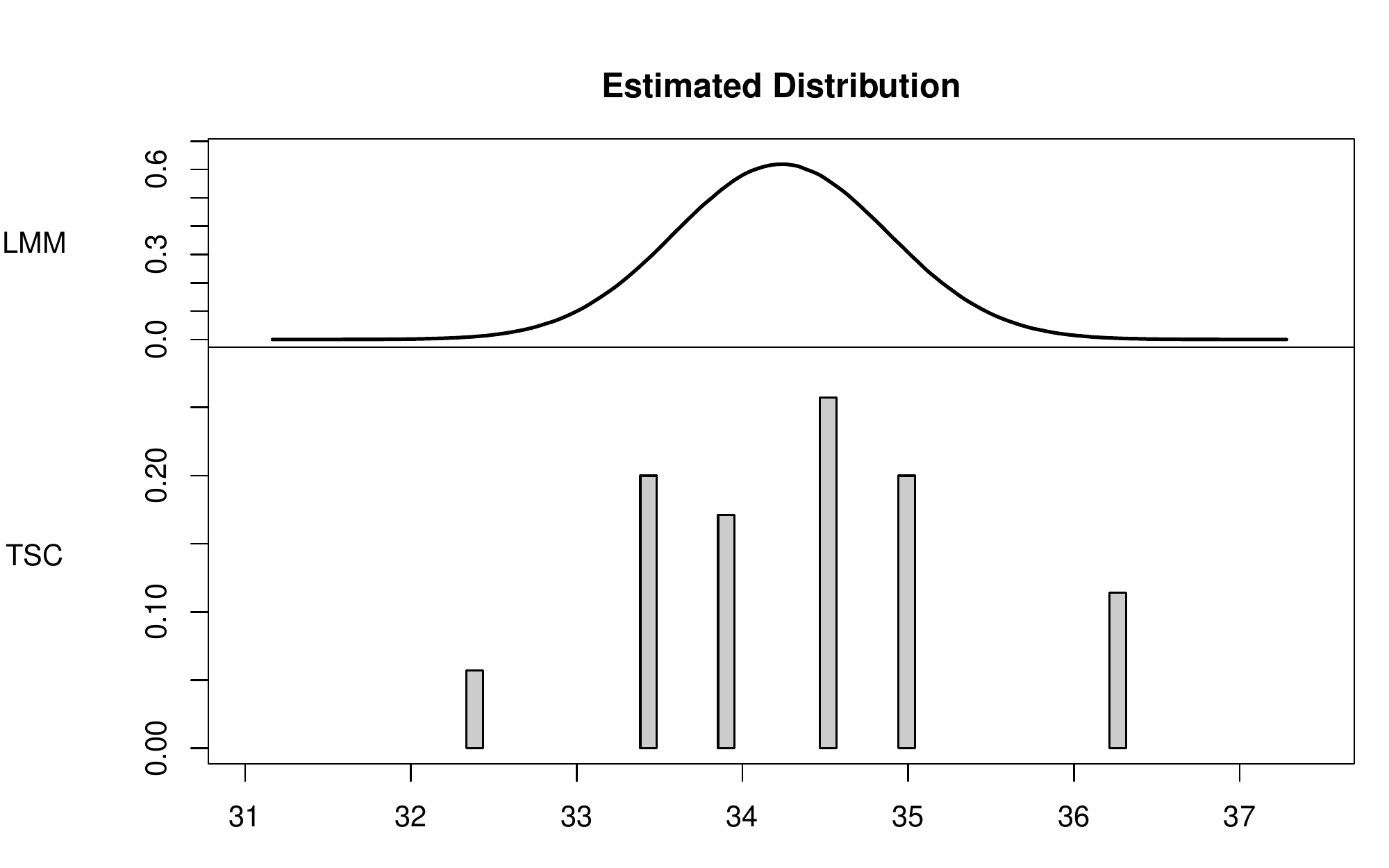}
\caption{Comparison of the estimated distribution of the mixed model and the school-specific intercepts of tree-structured clustering (CTB data).}
\label{fig:ctb_comparison}
\end{figure}

\begin{table}[!ht]
\caption{Estimation results of the illustrative example (CTB data) using the classical mixed model, tree-structured clustering and the finite mixture model.}
\begin{center}
\begin{scriptsize}
\begin{tabularx}{1\textwidth}{Xcccccc}
\toprule
\bf{Predictor}&\multicolumn{2}{c}{\bf{LMM}}&\multicolumn{2}{c}{\bf{TSC}}&\multicolumn{2}{c}{\bf{FIN}}\\
&Coefficient&$95\%$-CI&Coefficient&$95\%$-CI&Coefficient&$95\%$-CI\\
\midrule
gender&\,-0.106&[\;-0.475,\;\;0.298]&-0.088&[\;-0.478,\;\;0.313]&-0.084&[\;-0.473,\;\;0.309]\\
\\
$\beta_0$&34.235&[33.964,34.542]&---&---&---&---\\
$\sigma^2_{\text{rand}}$&\;\,0.416&[\;\;0.394,\;\;1.353]&---&---&---&---\\
\bottomrule
\end{tabularx}

\vspace{0.2cm}

\begin{tabularx}{1\textwidth}{Xcccc}
\toprule
\bf{School-specific intercept}&\multicolumn{2}{c}{\bf{TSC}}&\multicolumn{2}{c}{\bf{FIN}}\\
&Cluster&Coefficient&Cluster&Coefficient\\
\midrule
$\beta_{i0}$&1,16&32.384&1,4,6,7,9,16,&33.508\\
&4,18,19,20,21,22,28&33.434&18,19,20,21,&\\
&6,7,9,{\color{gray}11,29},30&33.904&22,28,30&\\
&{\color{gray}3,5,12,14,15,25,26,31,34}&34.517&{\color{gray}2,3,5,8,10,11,12,13,14,}&34.689\\
&{\color{gray}2,10,13,17,23,24,32}&34.999&{\color{gray}15,17,23,24,25,26,27,}&\\
&{\color{gray}8,27,33,35}&36.264&{\color{gray}29,31,32,33,34,35}&\\
\bottomrule
\end{tabularx}
\end{scriptsize}
\end{center}
\label{tab:app_ctb_results}
\end{table}

\section{An Illustrative Example} \label{sec:example}

Before giving details how to grow trees and estimate the proposed model \eqref{eq:fixedeffects_m0} we want to illustrate the procedure by use of an application. We consider a data set from CTB/McGraw-Hill, a division of the Data Recognition Corporation (DRC). For a description of the original data, see \citet{de2004explanatory}. The data includes results of an achievement test that measures different objectives and subskills of subjects in mathematics and science. For our investigation we use the results of 1500 grade 8 students from 35 schools. They had to respond to 56 multiple-choice items (31 mathematics, 25 science). The response $y_{ij}$  is the overall test score of student $j$ in school $i$, defined as the number of correctly solved items. The main objective is to adequately describe the heterogeneity of the 35 schools. As additional covariate we include the gender  of the students (male: 0, female: 1). The summary statistics of the test scores and the covariate gender is given in Table \ref{tab:app_ctb_summary}.
By using the proposed tree-structured approach the model that was obtained has the form
\[
\mu_{ij}=\beta_G \cdot G_{ij} + \sum_{k=1}^{m_0}\beta_{i0}^{(k)}I(i \in S_{k0}),\quad i=1,\hdots,35,
\]
where $G_{ij} \in \{0,1\}$ denotes the gender of student $j$ in school $i$, $S_{10},\hdots,S_{m_00}$ is a partition of the 35 schools and $\beta_{i0}^{(k)},k=1,\hdots,m_0$, denote the effects of the corresponding clusters.

The coefficient paths of the school-specific intercepts obtained by tree-structured clustering are shown in Figure \ref{fig:results_ctb}. The coefficient paths build a tree that successively partitions the schools in terms of the performance of students. The left end refers to the global intercept estimated as an average over the 35 schools. On the right end of the coefficients paths all possible splits have been performed and the estimated coefficients correspond to those of a simple fixed effects model without clustering. The optimal number of splits that is selected by the algorithm, is marked by the dashed line. It is seen that estimates change strongly in the first steps, but after about ten splits the estimates are very stable.

A graphical comparison of the estimated normal distribution of the random effects using a classical linear mixed model and the distribution of the school-specific intercepts of the tree-structured model is shown in Figure \ref{fig:ctb_comparison}. It illustrates the main advantage of the tree-structured model. There is no distributional assumption on the school-specific intercepts, especially no assumption of symmetry. The number of schools in each cluster are quite different and not symmetric. Clustering of similar schools strongly reduces the complexity of the fixed effects model and makes interpretation of school-specific differences very easy. There are two small clusters of schools where the performance in the test considerably deviates upwards or downwards,  the differences between the clusters with medium performance are smaller.

Table \ref{tab:app_ctb_results} shows an overview of the estimation results obtained by using the classical linear mixed model (LMM), the proposed tree-structured model (TSC) and a finite mixture model (FIN), where only the intercepts are allowed to vary over the components.
Confidence intervals are obtained by using bootstrap procedures, where the model is fitted repeatedly on sub samples of size $n$ that are obtained by drawing with replacement. The results here are obtained by $2000$ sub samples. It is seen that all of the methods did not find a significant effect for covariate gender. The performance of males and females seems not to differ systematically. The variance obtained by the mixed model is significantly different from zero, which suggests that  heterogeneity of schools is definitely present. The lower panel in Table \ref{tab:app_ctb_results} shows the estimated partition of schools obtained by the tree-structured model and the finite mixture model. In the latter case, model selection by AIC and BIC both yield the same result.
Tree-structured clustering identifies six clusters of schools until further splits are no longer significant (for details of the algorithm see Section \ref{sec:fitting procedure}).
The finite mixture approach identifies only two clusters of schools. This illustrates the tendency of the finite mixture approach to find a small number of clusters, which will be investigated later. For comparison in Table \ref{tab:app_ctb_results} the schools that belong to the two clusters found by the finite mixture model are coloured in black and grey.

\section{Fitting procedure} \label{sec:fitting procedure}

In this section we give details of the algorithm that yields the tree-structured model.
Let us again consider the model with unit-specific intercepts after the first split, which has the form
\begin{equation} \label{eq:fixedeffects_two2_fitting}
\eta_{ij}=\xb_{ij}^\top\betab+\beta_{i0}^{(1)}I(i\in S_{10})+\beta_{i0}^{(2)}I(i\in S_{20}).
\end{equation}
When determining the first split for the nominal predictor $i \in \{1,\hdots,n\}$ one has to consider all possible partitions of the two subsets $S_{10}$ and $S_{20}$. Altogether there are $2^{n-1}-1$ possible splits, which can be a very large number. It has been shown in earlier research that it is not necessary to consider all possible partitions, see \citet{BreiFrieOls:84} and \citet{Ripley:96} for binary outcomes and \citet{fisher1958grouping} for quantitative outcomes. It is sufficient to order the predictor categories, here the measurement units, with respect to the means of the response and to treat the predictor as if the categories were ordered. In a first step, units are ordered according to their maximum-likelihood estimates, so that $\hat{\beta}_{(10)}\leq\hat{\beta}_{(20)}\leq\hdots\leq\hat{\beta}_{(n0)}$. Then one considers splits of adjacent measurement units to obtain the optimal split. % with regard to various split measures.
To use this simplification one starts with  an equivalent representation of model \eqref{eq:fixedeffects_two2_fitting} given by
\begin{equation*} \label{eq:onesplit_one}
\eta_{ij}=\xb_{ij}^\top\betab+\beta_{0}+\alpha_{i0} I(i > c),
\end{equation*}
with $\beta_{i0}^{(1)}=\beta_{0}$ and $\beta_{i0}^{(2)}=\beta_{0}+\alpha_{i0}$. The set $C$ of possible thresholds $c$ is from $\{1,\hdots,n-1\}$. The fitting procedure considered in the following uses this model as building block.  By iterative splitting of adjacent measurement units  the searched-for clustering is obtained.

\subsubsection*{Basic Algorithm}

The basic algorithm for the model with unit-specific intercept is the following.

\vspace{0.5 cm}
\hrule
\begin{center}{\bf Tree-Structured Clustering -- Unit-specific intercept}\end{center}

\begin{description}
\item{\it Step 1 (Initialization)}

\begin{itemize}
\item[(a)] Estimation: Fit the candidate GLMs with predictors
\[
\eta_{ij} = \xb_{ij}^\top\betab+\beta_{0}+ \alpha_{i0} I(i > c_{i0}), \quad i=1,\dots,n-1
\]
\item[(b)] Selection

Select the model that has the best fit. Let $c_{i_10}^*$ denote the best split.
\end{itemize}

\item{\it Step 2 (Iteration)}

For $\ell=1,2,\dots$,

\begin{itemize}
\item[(a)] Estimation: Fit the candidate models with predictors
\[
\eta_{ij} = \xb_{ij}^\top\betab+\beta_{0}+ \sum_{s=1}^\ell\alpha_{i_s0} I(i > c_{i_s0}^*) + \alpha_{i0} I(i > c_{i0}),
\]
for all values $c_{i0} \in C \setminus \{c_{i_10}^*,\hdots,c_{i_\ell0}^*\}$

\item[(b)] Selection

Select the model that has the best fit yielding the split point $c_{i_{\ell+1}0}^*$.
\end{itemize}
\end{description}
\hrule
\vspace{0.5 cm}
In each selection step of the algorithm one has to identify the best split and during the iterations one has to decide when to stop. Common splitting criteria for tree-based methods are impurity measures that have already been introduced by \citet{BreiFrieOls:84}. An alternative is to use a test statistic to evaluate which split most improves the explanatory power of the predictors. We will draw on the latter concept and use a procedure that is strongly related to the conditional inference framework proposed by \citet{Hotetal:2006}.

In each iteration one examines the null hypotheses $H_0:\alpha_{i0}=0$ for all remaining possible split points. This can, for example, be tested by a likelihood-ratio test. To determine the best split we simultaneously consider all test statistics $T_{i0}$ from the set of possible splits $c_{i0}$ and choose the split point for which $T_{i0}$ had the largest value. This corresponds to choosing the split with the smallest $p$-value obtained from the chi-squared distribution of the test statistic.
To determine the optimal number of splits  our strategy is to check if
the heterogeneity of measurement units is already modelled sufficiently in each step. Before executing one further split one tests the global null hypothesis that the current model completely captures the heterogeneity of the data against the alternative that the data is more heterogeneous. To decide for the first split one has to examine the null hypothesis $H_0:\beta_{10}=\beta_{20}=\hdots=\beta_{n0}$, which corresponds to the case of no heterogeneity. The hypothesis is tested by a likelihood-ratio test with significance level $\alpha$ and $n-1$ degrees of freedom, because $n-1$ differences of parameters are tested. Depending on the significance of this global test the selected split or no splitting is performed. After several splits only differences of units within already built clusters are tested. In the $\ell-th$ step $n-\ell$ differences have to be tested because $\ell-1$ splits are already performed. If a significant effect is found the selected split is performed, otherwise splitting is stopped. The proposed stopping criterion leads to a clear separation of the selection of splits and the splitting decision. In particular the splitting decision is not influenced by the previously identified ordering of measurement units.

%Since each likelihood ratio test statistic asymptotically follows a chi-squared distribution, one additionally obtains a $p$-value associated with the test statistic $T_{0i}$ of the selected split. Splitting is stopped if the $p$-value exceeds a certain pre-specified threshold. In each step one should take into account the number of possible splits and adapt for multiple testing errors. Given overall significance level $\alpha$ one simply uses the Bonferroni procedure and stops if $p_\ell>\alpha/(n-(\ell-1))$ because in the $\ell-th$ iteration there are $n-(\ell-1)$ possible splits. Thus, the overall error rate is under control.

The result of the fitting procedure is a sequence of $m_0-1$ selected split points $c_{i_10}^*,\hdots,c_{i_{m_0-1}0}^*$ and corresponding parameter estimates $\hat{\alpha}_{i_10},\hdots,\hat{\alpha}_{i_{m_0-1}0}$. Ordering of the selected split points
%, so that $c_{(i0_1)}^* < c_{(i0_2)}^* < \hdots < c_{(i0_{m_0-1})}^*$,
yields the desired clustering of ordered units $\{1,\hdots,c_{(i_10)}^*\}$, $\{c_{(i_10)}^*+1,\hdots,c_{(i_20)}^*\}$, $\hdots$ , $\{c_{(i_{m_0-1}0)}^*+1,\hdots,n\}$.
%Given ordered units one obtains the clustering after ordering the selected split points, so that $c_{(i0_1)}^* < c_{(i0_2)}^* < \hdots < c_{(i0_{m_0-1})}^*$.
The corresponding intercepts $\beta_{i0}^{(k)}$ for each cluster are then given by
\begin{equation*}
\hat{\beta}_{i0}^{(k)}=\hat{\beta}_{0}+\sum_{s=1}^{k-1}{\hat{\alpha}_{(i_s0)}},\quad k=1,\hdots,m_0.
\end{equation*}
During the iterations only the selected split points but no estimates from previous steps are kept. All coefficients of the models, including the parameters $\betab$ of the linear term, are refitted in each step and the final estimates are those from the last iteration.

\section{Related Approaches} \label{sec:related}

In the following we will briefly consider alternative methods that account for unobserved heterogeneity and are  related to our tree-structured model.
One of the approaches is a competitor to the method proposed here and will also be included in the simulations.
%There are two approaches that also may be used to identify clusters of units that differ with regard to their intercepts. As they are competing approaches we will briefly sketch them and include them in our simulations in Section \ref{sec:simulations}.

Clustering of units can also be obtained by penalized maximum likelihood estimation as proposed more recently by \citet{TuOelkFixed}. Let $\betab_0^T=(\beta_{10},\dots,\beta_{n0})$  denote the intercepts of the fixed effects model. An estimation procedure that identifies clusters is obtained by maximizing the penalized log-likelihood $l_p(\betab,\betab_0)=l(\betab,\betab_0)-\lambda J(\betab,\betab_0)$, where $l(\betab,\betab_0)$ denotes the unpenalized log-likelihood, $J(\betab,\betab_0)$ is a specific penalty term and $\lambda$ is a tuning parameter. The penalty term that enforces clustering of unit-specific intercepts is given by
\begin{equation*}
J(\betab,\betab_0)=\sum_{r>s}{\left|\beta_{r0}-\beta_{s0}\right|},
\end{equation*}
where only pairwise differences of the unit-specific intercepts are included. If $\lambda=0$, one obtains the unpenalized maximum-likelihood estimates and each unit has his own intercept. If $\lambda \rightarrow \infty$, all units are fused to one cluster with the same intercept. For a comparison we use the corresponding R-package \texttt{gvcm.cat} proposed by \citet{gvcm.cat:2015} in our simulations. The use of such penalties in ANOVA was already proposed by \citet{BonRei:2008} and for variable selection by \citet{GerTut:2009a} and \citet{TutGer:2014}. A problem with the method is that the penalty contains $n(n-1)/2$ differences and therefore the  algorithm becomes extremely demanding for large values of $n$. It typically fails if the number of groups is larger than 50 or 60.

The method proposed here should be distinguished from  the  mixed effects regression trees (MERT) proposed by \citet{hajjem2011} and the RE-EM trees, which were independently proposed by \citet{sela2012re}. The basic concept is to combine a linear mixed effects model for clustered data and a standard regression tree. The substantial difference is that the tree is not applied to the random or unit-specific effects of the model but to the fixed effects term. The predictor of the estimated model has the form $\eta_{ij}=f(\xb_{ij})+\zb_{ij}^\top\bb_i$, where $\bb_i\sim N(0,\Sigmab_{rand})$. It is the  function $f(\xb_{ij})$ that is estimated by a standard regression tree. The model yields random effects that are node-invariant and therefore does not focus on the similarity of units but rather on the dissimilarity of observations within units.

An alternative Bayesian approach to model clustered random effects is based on Dirichlet processes. Dirichlet processes were proposed by \citet{Ferguson:73} and studied, for example, by  \citet{Sethuraman:94} and \citet{hjort2010bayesian}. The main advantage of Dirichlet processes is their  cluster property, which allows to flexibly model discrete distributions.  Assuming a  Dirichlet process for the distribution of random effects creates ties among the random effects. The resulting Dirichlet process mixture yields clusters of units. Dirichlet process priors have been used  within the linear mixed model framework by \citet{bush1996semiparametric} and \citet{muller1997bayesian}. A frequentist approach to linear mixed models with Dirichlet process mixtures was given by
\citet{heinzl2013clustering}, a combination of Dirichlet processes and fusion penalties was considered in \citet{heinzl2014clustering}, \citet{heinzl2015additive}. The approach works for linear models, but extensions to generalized mixed models seem not available.

\section{Simulations} \label{sec:simulations}

In the following we investigate the performance of the proposed tree-structured model and compare it to competing methods. The focus is on data settings with clusters of units that share the same effect on the response and where the strict assumptions of the mixed model do not hold. We are in particular interested in the estimation accuracy and the clustering performance. We will compare the generalized fixed effects model (GFM), the generalized mixed model (GMM), the tree-structured model (TSC), the model based on penalized maximum-likelihood estimation (PENL), the finite mixture model with model selection by AIC (FINA) and the finite mixture model with model selection by BIC (FINB).

We consider several simulation scenarios where the overall number of observations is 800, made up of the components $n=200/n_i=4$, $n=100/n_i=8$, $n=40/n_i=20$ or $n=20/n_i=40$. In addition to the unit-specific intercepts we include one continuous covariate $x_1$ with $x_{ij1} \sim N(0,1)$ and one binary covariate $x_2$ with $x_{ij2} \sim B(1,0.5)$. Unit-specific intercepts $\beta_{i0}$ are drawn symmetrically from a normal distribution or are drawn from a chi-square distribution that is skewed. In order to obtain clusters of units, the intercepts are sorted according to size and divided into balanced groups. The average over the intercepts of each group is defined as the new unit-specific intercept $\beta_{i0}^{(k)},\,k=1,\hdots,m_0$. We consider scenarios with $m_0\in\{5,10\}$. Therefore, the true simulated size of clusters varies between 2 for the scenarios with $n=20$, $m_0=10$, and 40 for the settings with $n=200$, $m_0=5$.

\subsubsection*{Correlation between Intercepts and Covariates}

An important assumption of the mixed model is that the unit-specific intercepts are independent from the predictors $\xb$. In order to break this assumption we simulate data with correlations $\rho=\text{corr}(\beta_{i0},x_{ij1}) \neq 0$.  For the simulation we use a sequential procedure adopted from \citet{TuOelkFixed}. Consider the case of normal distributed intercepts $\beta_{i0}$. Here, values are first generated by $\beta_{i0}\sim N(\mu_b,\sigma_b^2)$ and $x_{ij1}\sim N(0,1)$. Afterwards $x_{ij1}$ is transformed according to the bivariate normal distribution of $(\beta_{i0}, x_{ij1})$ with the corresponding correlation. We consider scenarios with $\rho\in\{0,0.8\}$. In the case of chi-squared distributed intercepts the joint distribution of $(\beta_{i0}, x_{ij1})$ is not bivariate normal, but we can use the same transformation for $x_{ij1}$ yielding the same empirical correlations.

\subsubsection*{Evaluation Criteria}

We compare the estimated coefficients to the true parameters by calculating mean squared errors (MSEs). We distinguish between the MSE of the unit-specific intercepts $\frac{1}{n}\sum_{i=1}^{n}{(\hat{\beta}_{i0}-\beta_{i0})^2}$, referred to as \textit{intercepts}, and the MSE of the effects of the two covariates $\frac{1}{2}\sum_{d=1}^{2}{(\hat{\beta}_d-\beta_d)^2}$, referred to as \textit{linear term}. Concerning the mixed model, coefficients $\hat{\beta}_{i0}$ are computed as the sum of the estimated posteriori modes and the fixed intercept $\hat{\beta}_0$. In addition the number of clusters determined by the different approaches are of interest. All the presented evaluations are based on 100 replications.

\begin{figure}[!ht]
\centering
\includegraphics[width=0.9\textwidth]{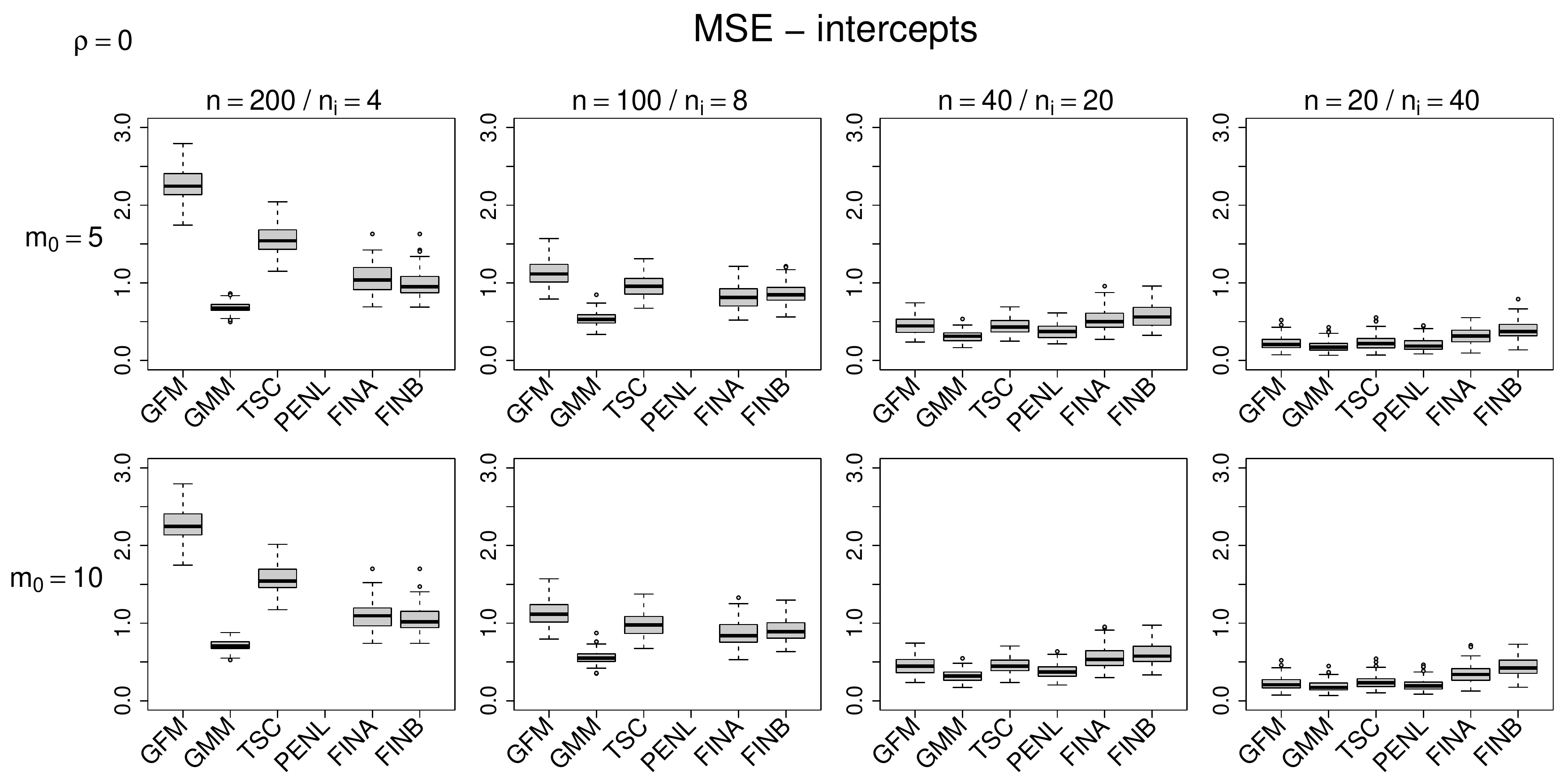}

\includegraphics[width=0.9\textwidth]{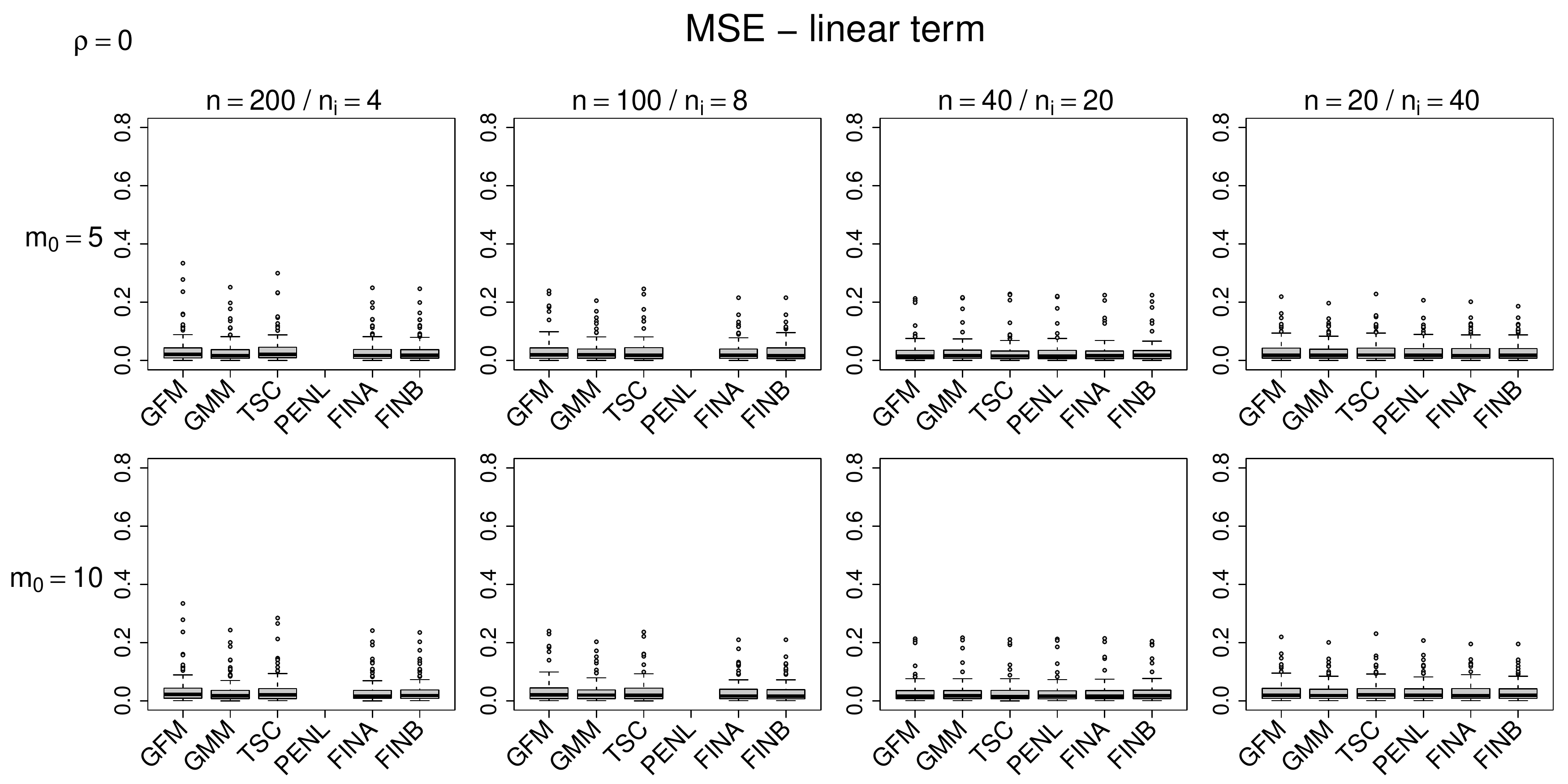}

\caption{MSEs of intercepts (upper panel) and the linear term (lower panel) for the settings with normal response, normal intercepts and $\rho=0$.}
\label{fig:sim_mse_nono0}
\end{figure}

\begin{figure}[!ht]
\centering
\includegraphics[width=0.9\textwidth]{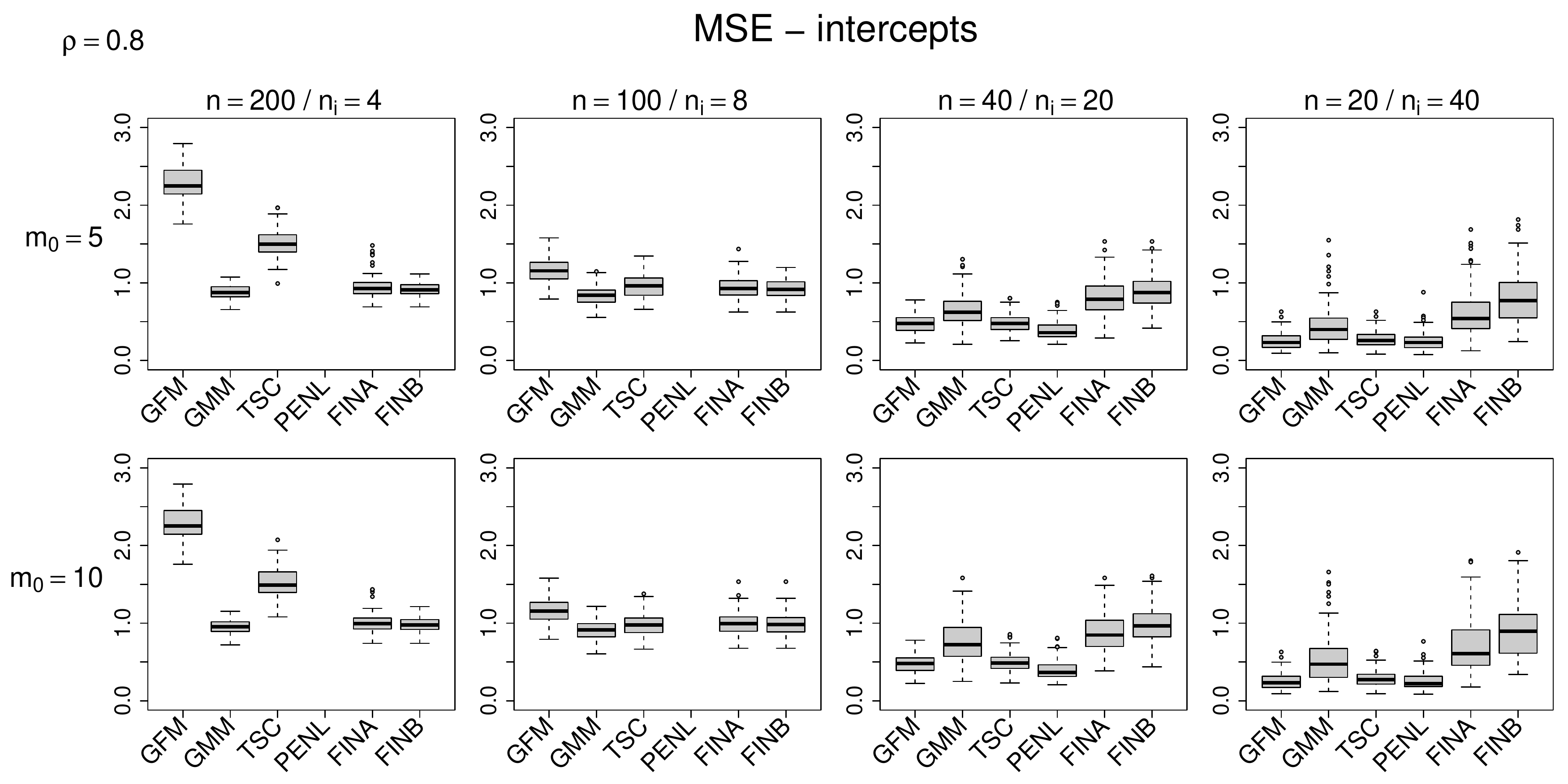}

\includegraphics[width=0.9\textwidth]{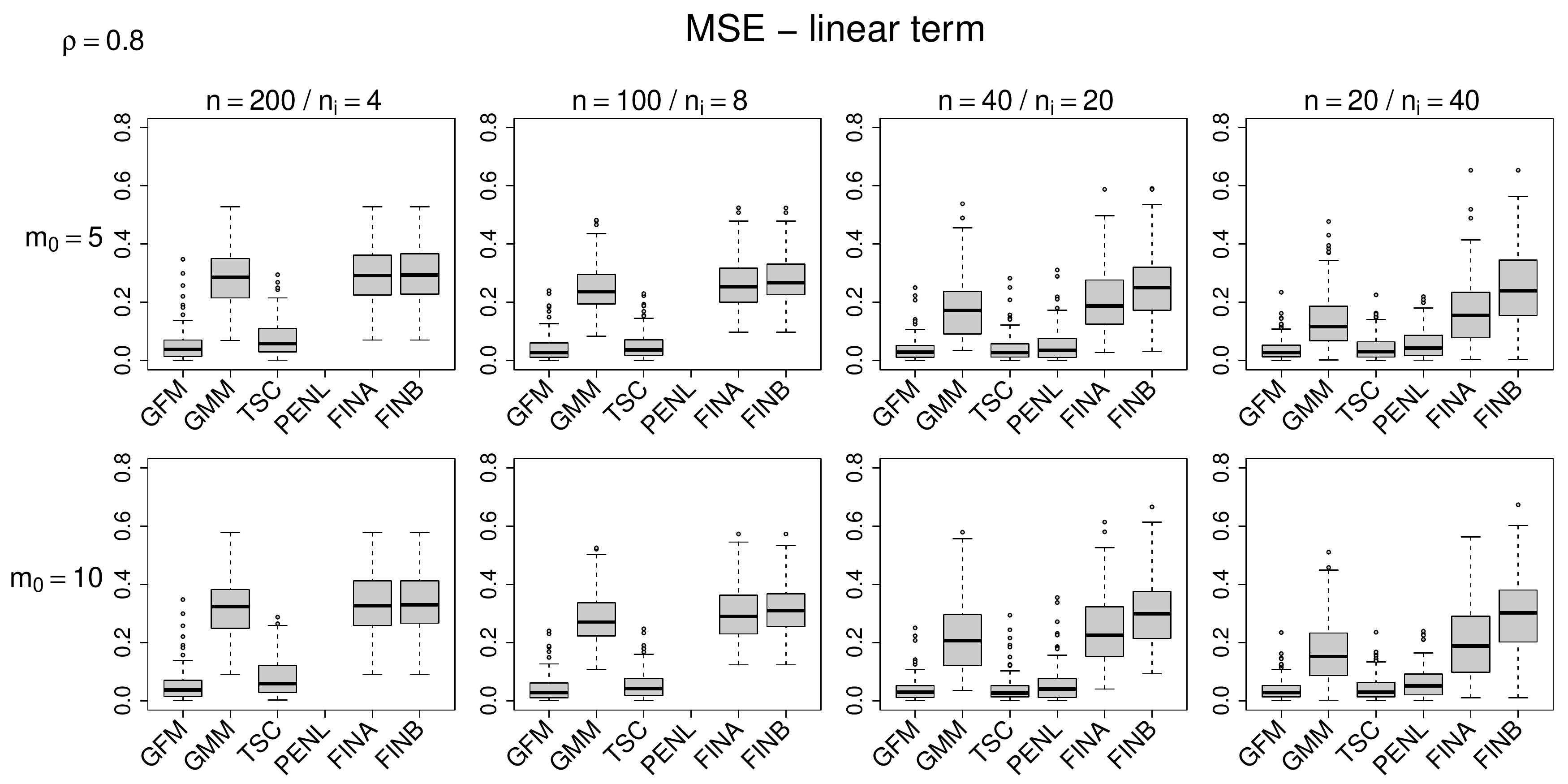}

\caption{MSEs of intercepts (upper panel) and the linear term (lower panel) for the settings with normal response, normal intercepts and $\rho=0.8$.}
\label{fig:sim_mse_nono8}
\end{figure}

\begin{figure}[!ht]
\centering
\includegraphics[width=0.9\textwidth]{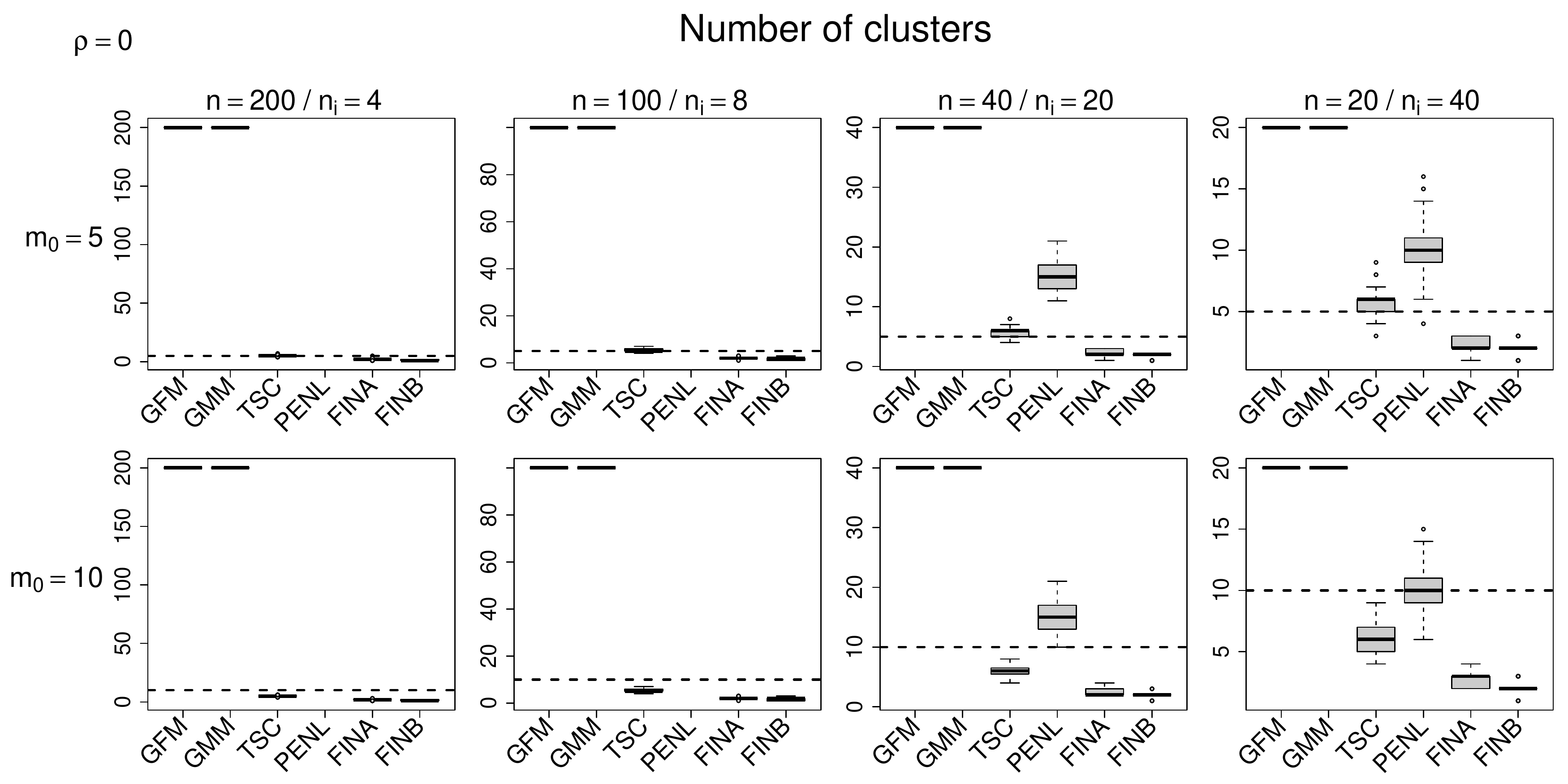}

\includegraphics[width=0.9\textwidth]{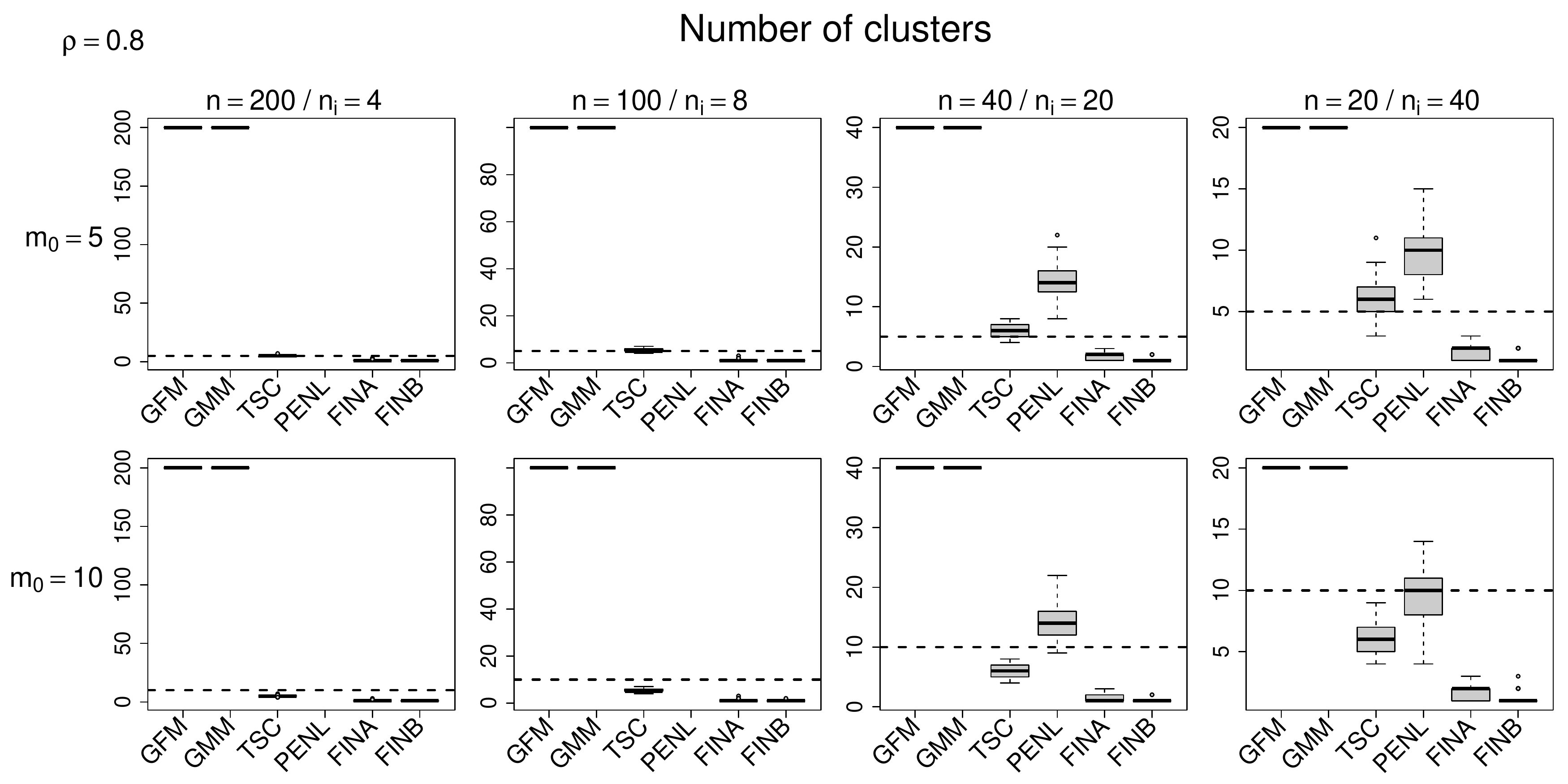}

\caption{Selected number of clusters for the settings with normal response, normal intercepts, $\rho=0$ (upper panel) and $\rho=0.8$ (lower panel). The true number of clusters $m_0$ is marked by dashed lines.}
\label{fig:sim_nc_nono}
\end{figure}

\begin{figure}[!ht]
\centering
\includegraphics[width=0.9\textwidth]{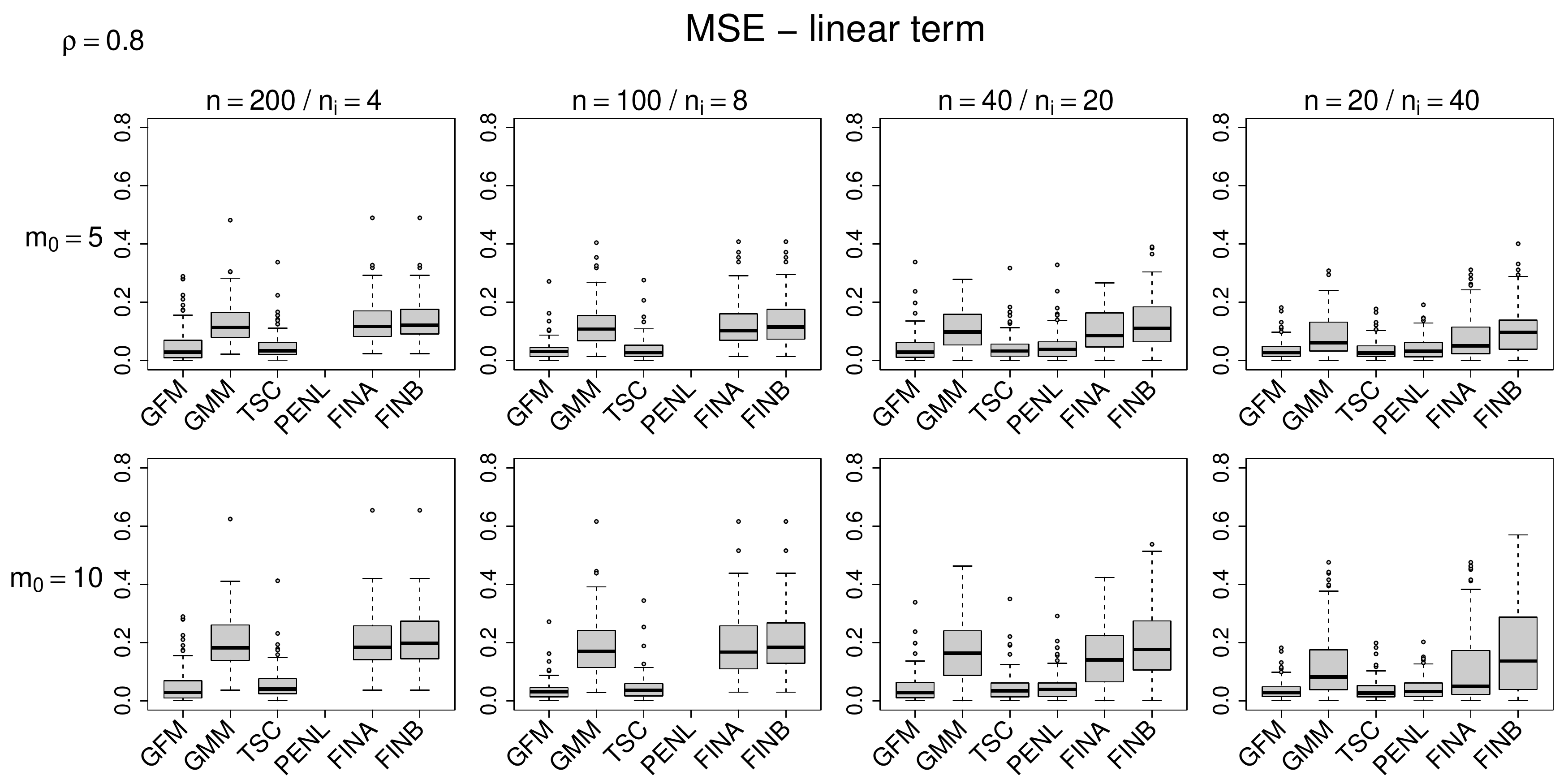}
\caption{MSEs of the linear term for the settings with normal response, chi-squared intercepts and $\rho=0.8$.}
\label{fig:sim_mse_nochi8}
\end{figure}

\begin{figure}[!ht]
\centering
\includegraphics[width=0.67\textwidth]{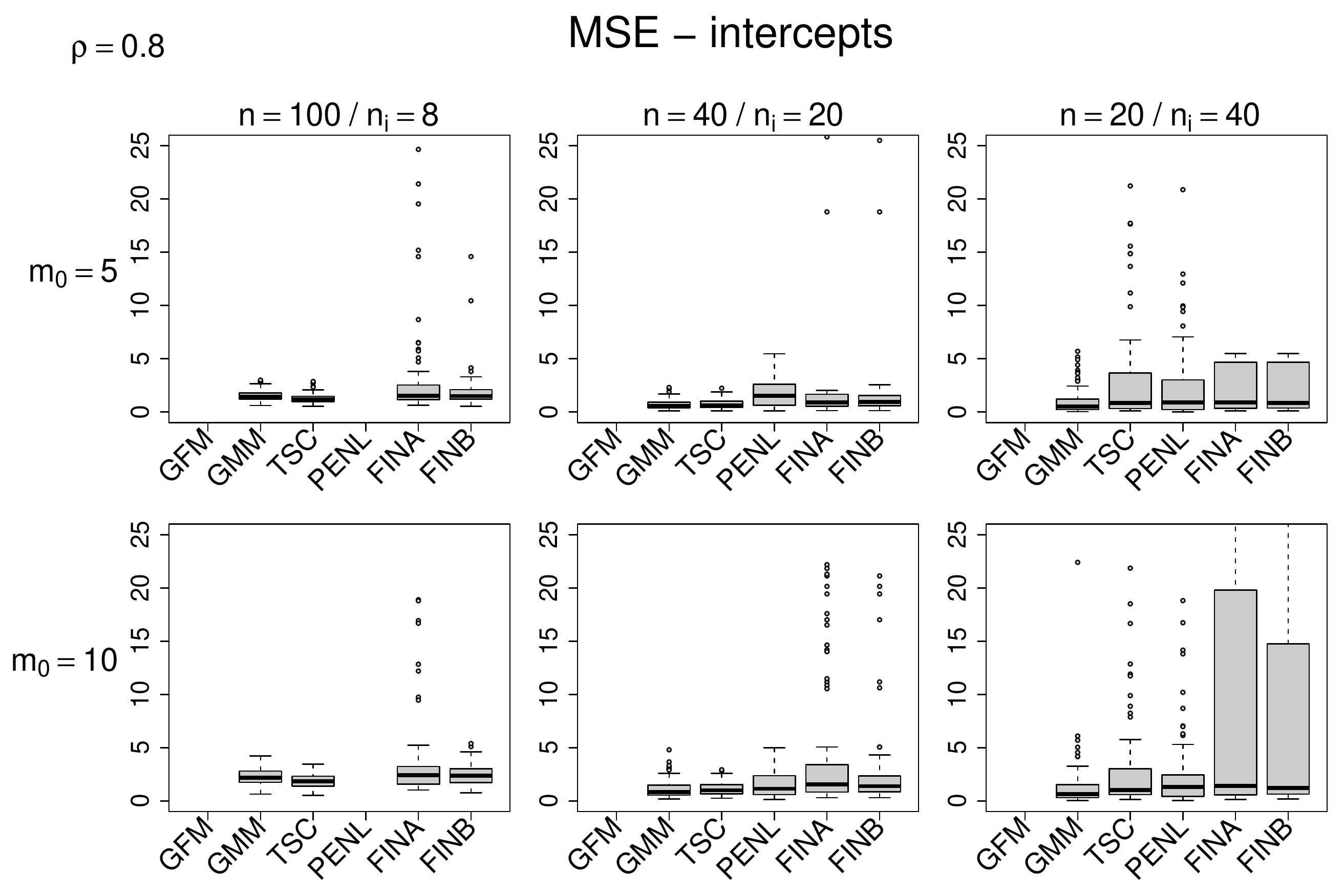}

\includegraphics[width=0.67\textwidth]{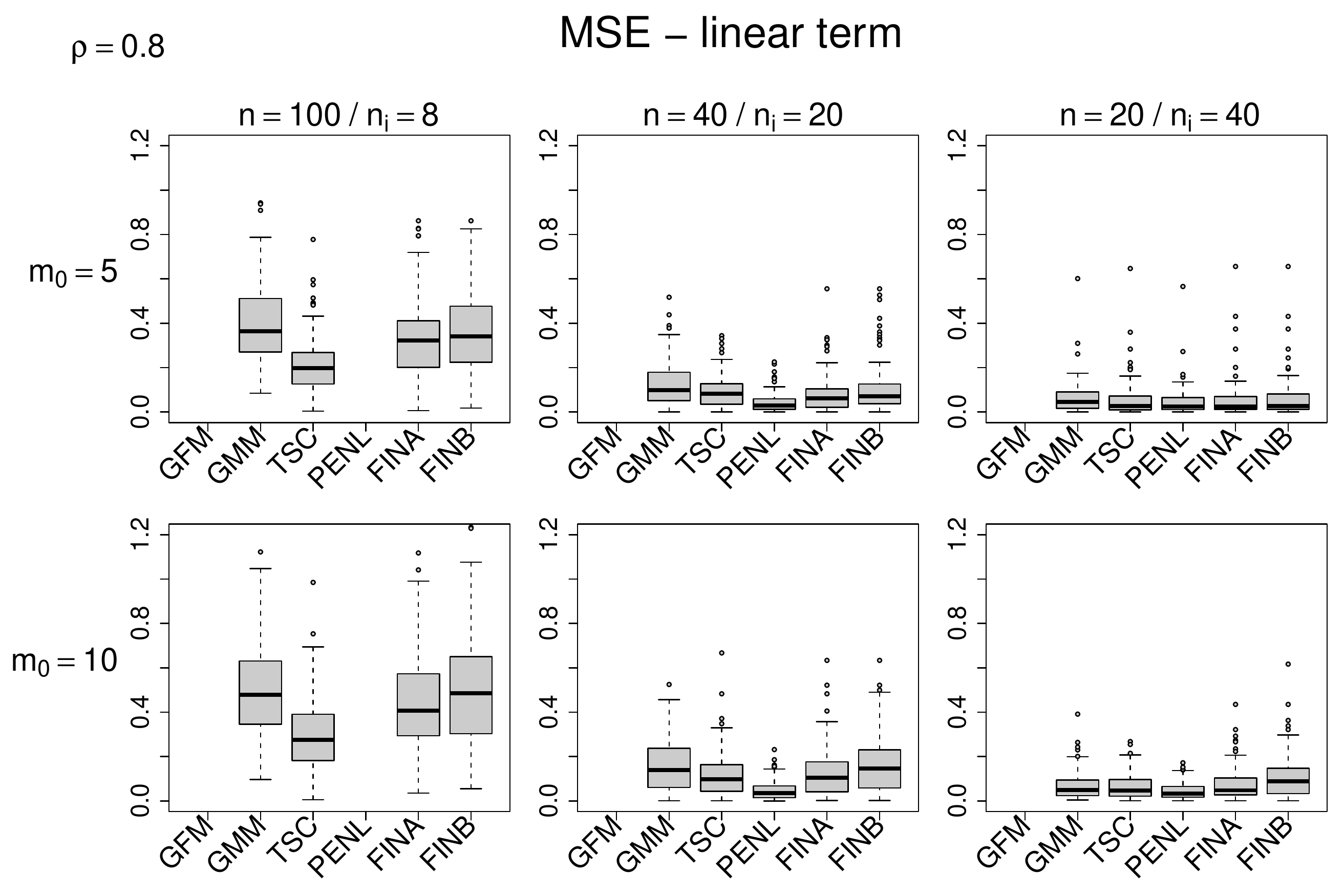}

\caption{MSEs of intercepts (upper panel) and the linear term (lower panel) for the settings with binary response, chi-squared intercepts and $\rho=0.8$.}
\label{fig:sim_mse_binchi8}
\end{figure}

\begin{figure}[!ht]
\centering
\includegraphics[width=0.67\textwidth]{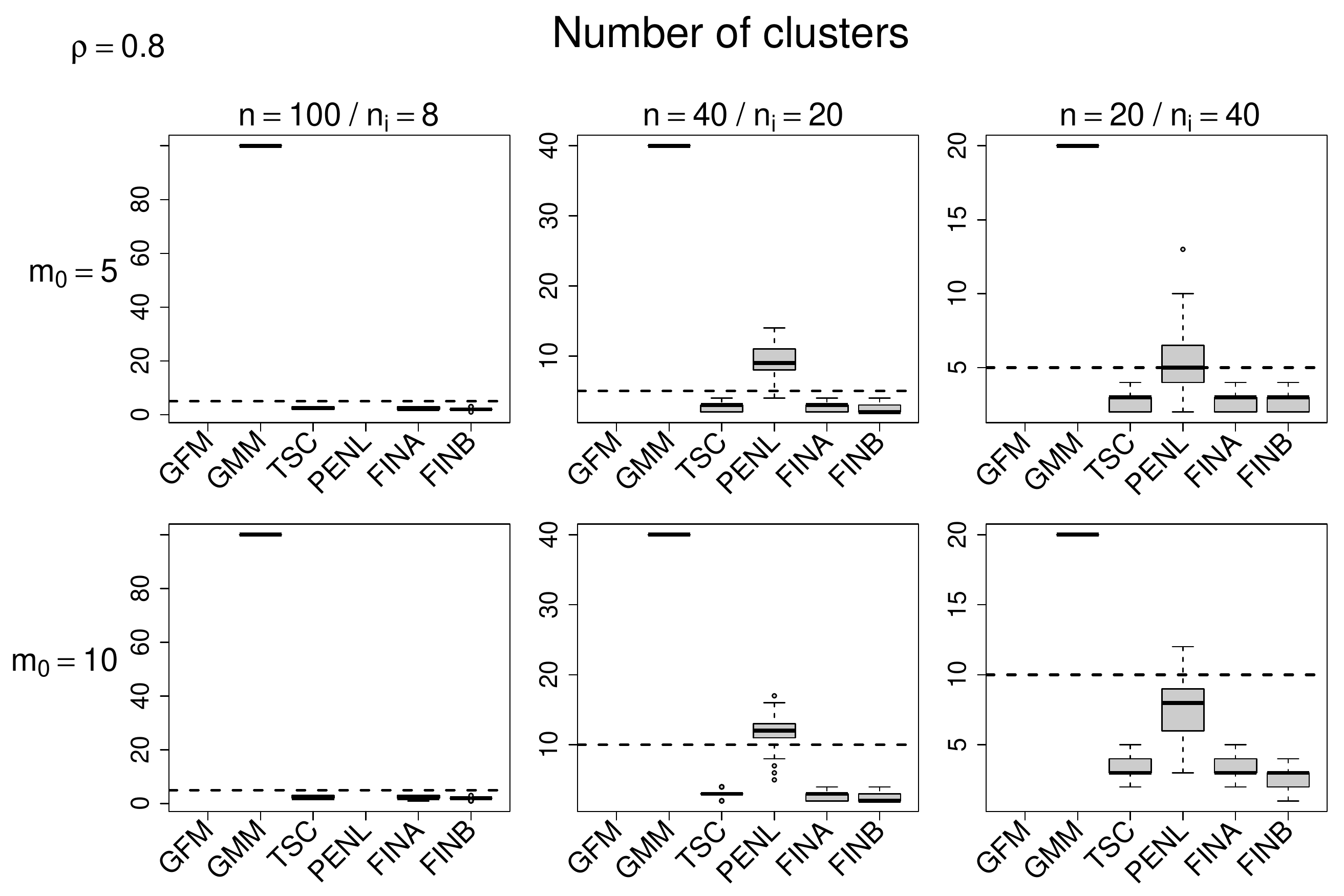}
\caption{Selected number of clusters for the settings with binary response, chi-squared intercepts and $\rho=0.8$. The true number of clusters $m_0$ is marked by dashed lines.}
\label{fig:sim_nc_binchi8}
\end{figure}

\subsection{Normal Response} \label{subsec:sim_normal}

We start with simulation scenarios where the responses $y_{ij},\,i=1,\hdots,n,\,j=1,\hdots,n_i$ are normally distributed with $\epsilon_{ij} \sim N(\mu_{\epsilon}=0,\sigma_{\epsilon}^2=3^2)$.
Here we set $\beta_1=\beta_2=2$ as the true parameters of the two covariates. In the first case we consider cluster-specific intercepts that were generated from the fusion of parameters that follow a standard normal distribution.  %$\beta_{i0}\sim N(\mu_b=0,\sigma_b^2=1^2)$.

%\subsubsection*{Effective Degrees of Freedom}

It is important to mention that in the above setting the effective number of parameters for the mixed model heavily depends on the variance $\sigma^2_{\epsilon}$ of the response and the variance $\sigma^2_b$ of the random intercepts. Following \citet{RupWanCar:2003}, the effective degrees of freedom for the random intercepts for a linear random intercept model are
\[
df_b=\frac{(n-1)n_i}{n_i+\frac{\sigma^2_{\epsilon}}{\sigma^2_b}}.
\]
If $\sigma^2_b \rightarrow 0$ or $\sigma^2_{\epsilon} \rightarrow \infty$ the result is a model with only one intercept and if $\sigma^2_b \rightarrow \infty$ or $\sigma^2_{\epsilon} \rightarrow 0$ the result is a model with $n$ intercepts, corresponding to the fixed effects model. With $\sigma^2_{\epsilon}=9$ and $\sigma^2_b=1$ one obtains the effective degrees of freedom $61.2$, $46.5$, $26.9$ and $15.5$ depending on the combination of parameters $n$ and $n_i$. Therefore, one is not too close to the fixed effects model, which allows a fair comparison of the mixed model and the tree-structured model.
%\\[1em]
In the second case with a skewed distribution for the unit-specific intercepts we use $\beta_{i0}\sim \chi^2(0.5)$ with $\sigma^2_b/2=0.5$ degrees of freedom. After centering of the coefficients one obtains the same empirical values $\mu_b=0$ and $\sigma_b^2=1$ as in the standard normal case.

Figure \ref{fig:sim_mse_nono0} shows the boxplots of the MSEs for the eight different settings generated by normally distributed intercepts and without correlation ($\rho=0$). As the approach by penalized likelihood estimation is computational unfeasible for a large number of units $n$, no results are displayed for the settings with $n=200$ and $n=100$. It is seen from the lower panel that all the approaches nearly show the same performance for the linear term. However, distinct differences are seen for the intercepts (upper panel). Although there are clusters of units the mixed model shows good performance for all settings. The fixed effects model performs poorly, especially for the settings with $n_i=4$, the finite mixture model performs poorly for the settings with $n=40$ and $n=20$. The estimates of the tree-structured model show better performance than the fixed effects model for smaller values of $n_i$ and comparable performance for larger values. The performance is the same as for the penalty approach if estimates exist.
The picture changes in   the  settings  with correlation $\rho=0.8$ between covariate $x_1$ and the unit-specific intercepts  (Figure \ref{fig:sim_mse_nono8}). For the linear term (lower panel) the performance of the mixed model and the finite mixture model  suffers strongly. In contrast, the estimation accuracy of the fixed effects model, the tree-structured model and the penalized likelihood approach is not affected by the correlation. In particular, the tree-structured model outperforms the penalty approach in all the settings in which the penalty approach works. The results for the intercepts (upper panel) do not change that much but  the mixed model and the finite mixture model is now competitive only for small values of $n_i$.

Boxplots of the selected number of clusters are given in Figure \ref{fig:sim_nc_nono} for $\rho=0$ (upper panel) and $\rho=0.8$ (lower panel). Since the fixed effects model and the mixed model do not build clusters of units, the given number of clusters for the two approaches is equal to the number of units. There are only minor differences between the settings with and without correlation. The number of clusters identified by the tree-structured model is very close to the true number for the settings with five clusters ($m_0=5$) but the true number of clusters is slightly underestimated in the settings with ten clusters. In contrast, the penalty approach selects a distinctly higher number of clusters with a strong variation. The finite mixture model consistently selects only too small number of clusters. On average only about two clusters are selected by AIC as well as by BIC.

The evaluations of the same settings with  cluster-specific intercepts that were generated by a chi-squared distribution  yield very similar results. In particular the performance of the mixed model seems not to be affected too strongly by the skewed distribution of the random intercepts. For illustration Figure \ref{fig:sim_mse_nochi8} shows the MSEs of the linear term for the settings with $\rho=0.8$. See the appendix for an overview of all results.

\subsection{Binary Response} \label{subsec:sim_bin}

In the following we briefly consider discrete response variables $y_{ij} \sim B(1,\pi_{ij})$, where $\pi_{ij}=\exp(\eta_{ij})/(1+\exp(\eta_{ij}))$. The structure of the simulated data sets remains the same but some modifications to the specifications in Section \ref{subsec:sim_normal} are necessary. The parameters of the linear term are set to $\beta_1=\beta_2=0.1$. For the cluster-specific intercepts we chose $\beta_{i0}\sim N(-0.8,2^2)$ or as skew counterpart $\beta_{i0}\sim \chi^2(2)$, centered such that $\mu_b=-0.8$. Since $n_i=4$ is a relatively small size when modelling   binary responses, we do not consider the corresponding settings. Furthermore, we omit the estimates of the fixed effects model because they are very unstable and often do not exist in this case. Accordingly, the order of measurement units used in the algorithm of the tree-structured model is not based on the estimates of the unrestricted model but by adding a small ridge penalty. 

In contrast to the settings with normal response, the results for the binary response as a whole seem to be more affected by a skewed distribution of the intercepts. In the following we will focus on the settings with chi-squared distributed intercepts and $\rho=0.8$, and refer to the appendix for further results.
Figure \ref{fig:sim_mse_binchi8} shows the MSEs of the unit-specific intercepts (upper panel) and the linear term (lower panel). Again the mixed model and the finite mixture model perform poorly with regard to the linear term, but there are only minor differences for $n=20$. Regarding the intercepts the average results are comparable for all the approaches. It is noticeable that one observes huge outliers for the finite mixture models, especially with model selection by AIC. It is most conspicuous for the settings with $n=20$, where the boxplots have been truncated.

The corresponding boxplots of the selected number of clusters are given in Figure \ref{fig:sim_nc_binchi8}. Here the tree-structured model only detects very few clusters (for $m_0=5$ and $m_0=10$) and is almost as restrictive as the finite mixture model. As before the penalty approach selects a higher number of clusters and has a stronger variation but the selected number of  clusters is  closer to the true number of clusters.

\begin{table}[!ht]
\caption{Description and distribution of the covariates used for the analysis of the Guatemala survey.}
\begin{center}
\begin{scriptsize}
\begin{tabular}{lllr}
\toprule
\bf{Variable}&\bf{Description}&\bf{Categories}&\bf{Frequency}\\
\midrule
\bf{ethn}&Mother's ethnicity&non-indigenous (Ladino)&612\\
&&indigenous, not speaking Spanish&286\\
&&indigenous, speaking Spanish&313\\
\midrule
\bf{momEd}&Mother's level of education&not finished primary&571\\
&&finished primary&607\\
&&finished secondary&33\\
\midrule
\bf{husEd}&Husband's level of education&not finished primary&430\\
&&finished primary&598\\
&&finished secondary&67\\
&&unknown&116\\
\midrule
\bf{husEmpl}&Husband's employment status&unskilled&45\\
&&professional&120\\
&&agricultural, self-employed&420\\
&&agricultural, employee&407\\
&&skilled service&219\\
\midrule
\bf{telev}&Frequency of TV usage&never&1034\\
&&not daily&52\\
&&daily&125\\
\midrule
\bf{momAge}&Mother $25$ years or older &no&583\\
&&yes&628\\
\midrule
\bf{toilet}&Modern toilet in house&no&112\\
&&yes&1099\\
\bottomrule
\end{tabular}
\end{scriptsize}
\end{center}
\label{tab:app_gua_summary}
\end{table}

\begin{table}[!ht]
\caption{Estimation results of the Guatemala survey using the generalized mixed model, tree-structured clustering and the finite mixture model.}
\begin{center}
\begin{scriptsize}
\begin{tabularx}{1\textwidth}{Xcccccc}
\toprule
\bf{Predictor}&\multicolumn{2}{c}{\bf{GMM}}&\multicolumn{2}{c}{\bf{TSC}}&\multicolumn{2}{c}{\bf{FIN}}\\
&Coefficient&95$\%$-CI&Coefficient&95$\%$-CI&Coefficient&95$\%$-CI\\
\midrule
\bf{ethn}&&&&&&\\
not spanish&-1.370&[-2.101,-0.774]&-1.090&[-2.469,-0.387]&-0.995&[-2.280,-0.556]\\
spanish&-0.720&[-1.235,-0.244]&-0.434&[-1.425, 0.005]&-0.335&[-1.338, 0.011]\\
\bf{momEd}&&&&&&\\
primary&\;0.645&[ 0.331, 1.048]&\;0.673&[ 0.298, 1.122]&\;0.646&[ 0.317, 1.078]\\
secondary&\;1.385&[ 0.303, 2.955]&\;1.405&[ 0.268, 3.046]&\;1.735&[ 0.364, 2.944]\\
\bf{husEd}&&&&&&\\
primary&\;0.785&[ 0.445, 1.236]&\;0.817&[ 0.437, 1.303]&\;0.843&[ 0.444, 1.301]\\
secondary&\;0.194&[-0.809, 1.186]&\;0.049&[-0.922, 1.286]&\;0.291&[-0.846, 1.311]\\
unknown&\;0.398&[-0.113, 0.951]&\;0.520&[-0.101, 1.006]&\;0.428&[-0.106, 0.962]\\
\bf{husEmpl}&&&&&&\\
professional&-0.210&[-1.150, 0.670]&-0.095&[-1.301, 0.820]&-0.408&[-1.336, 0.667]\\
agricult, self&-0.119&[-0.975, 0.721]&-0.065&[-1.044, 0.798]&-0.266&[-1.065, 0.716]\\
agricult, empl&-0.158&[-1.024, 0.656]&-0.100&[-1.092, 0.750]&-0.238&[-1.103, 0.723]\\
skilled&-0.199&[-1.079, 0.606]&-0.125&[-1.123, 0.661]&-0.300&[-1.134, 0.607]\\
\bf{telev}&&&&&&\\
not daily&\;0.355&[-0.497, 1.292]&\;0.226&[-0.601, 1.286]&\;0.241&[-0.548, 1.283]\\
daily&\;0.867&[ 0.312, 1.560]&\;0.928&[ 0.290, 1.570]&\;0.735&[ 0.307, 1.524]\\
\\
momAge&\;0.099&[-0.208, 0.403]&\;0.061&[-0.241, 0.411]&\;0.061&[-0.219, 0.401]\\
toilet&-0.869&[-1.833,-0.055]&-1.008&[-1.875, 0.092]&-0.839&[-1.808,-0.154]\\
\midrule
$\beta_0$&-0.011&[-1.223, 1.166]&---&---&---&---\\
$\sigma^2_{\text{rand}}$&\;1.250&[ 1.233, 2.416]&---&---&---&---\\
\bottomrule
\end{tabularx}

\vspace{0.2cm}

\begin{tabularx}{1\textwidth}{Xcccccc}
\toprule
\bf{Community-specific intercept}&\multicolumn{3}{c}{\bf{TSC}}&\multicolumn{3}{c}{\bf{FIN}}\\
&Cluster&Size&Coefficient&Cluster&Size&Coefficient\\
\midrule
$\beta_{i0}$&1&15&-1.286&1&33&-0.696\\
&2&17&-0.214&2&12&\;1.465\\
&3&13&\;1.448&&&\\
\bottomrule
\end{tabularx}
\end{scriptsize}
\end{center}
\label{tab:app_gua_results}
\end{table}

\section{A Further Application} \label{sec:application}

As second application we consider data derived from the National Survey of Maternal and Child Health in Guatemala in 1987. The data is available from the R-package \texttt{mlmRev} \citep{mlmRev:2014} and was also analysed by \citet{RodGold:2001}. The data contains observations of children that were born in the 5-year period before the survey. In our analysis we include 1211 children living in 45 communities. One observes a minimal number of 20, a maximal number of 50 and an average number of 26.9 pregnancies per community. The response $y_{ij}$  is  a binary outcome with $y_{ij}=0$ for traditional prenatal care and $y_{ij}=1$ for modern prenatal care, for example by doctors or nurses. The response is modelled by a logistic regression model logit$\left(P(y_{ij}=1)\right)=\eta_{ij}$. The heterogeneity of communities is modelled by the alternative approaches considered here. In total there are 733 pregnancies with traditional and 478 observed pregnancies with modern prenatal care.
The two binary and five categorical explanatory variables that characterize the children's mothers and their families are given in Table \ref{tab:app_gua_summary}.

An overview of the estimated coefficients when using a generalized mixed model (GMM), tree-structured clustering (TSC) and a finite mixture model (FIN) is given in Table \ref{tab:app_gua_results}. The $95\%$ confidence intervals were obtained by 2000 bootstrap samples. It can be seen from the results that the age of the mother at the time of the survey as well as the employment status of the husband do not have a significant effect on the form of prenatal care. The educational level of the mother as well as of the husband, however, have a strong impact. For births where the mother at least finished primary or the husband finished primary modern prenatal care was provided more likely compared to births of parents without any graduation. Indigenous mothers (speaking and not speaking Spanish) are also more likely to use traditional prenatal care than non-indigenous mothers. The existence of a modern toilet in the household does not favour the use of modern prenatal care, whereas it is preferred by families using the television regularly.

A comparison of the estimates obtained by the three methods does not show strong distinctions and no clear tendency. Differences occur for variable ethnicity (first rows in Table \ref{tab:app_gua_results}), for which the two estimates of the mixed model are larger than for TSC and FIN and for mothers that finished secondary (fourth row) for which the estimate of the finite mixture model is larger than for TSC and GMM.

\begin{figure}[!t]
\centering
\includegraphics[width=0.7\textwidth]{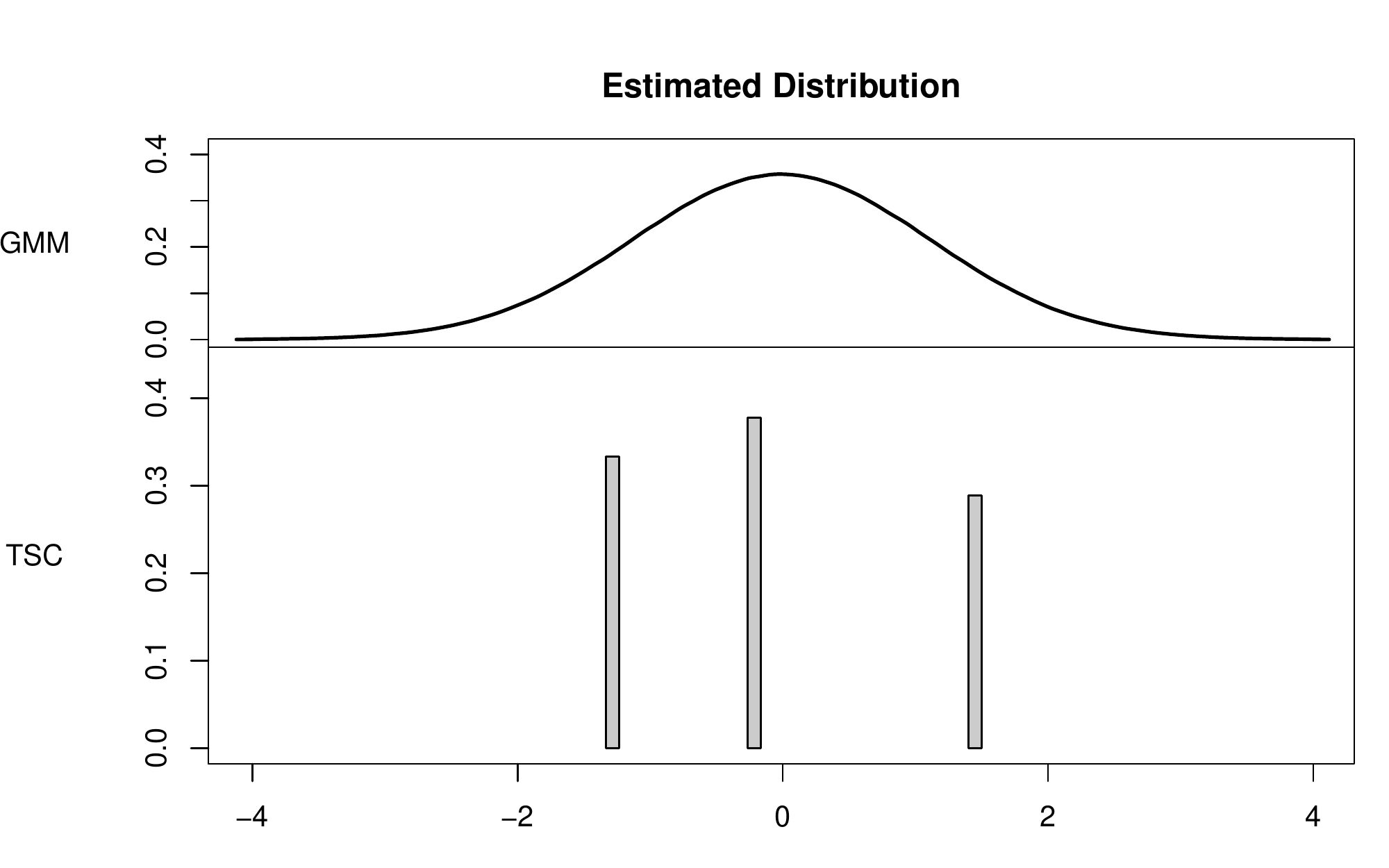}
\caption{Comparison of the estimated distribution of the mixed model and the community-specific intercepts of tree-structured clustering (Guatemala survey).}
\label{fig:gua_comparison}
\end{figure}

The estimated community-specific intercepts obtained by tree-structured clustering and the finite mixture model are given in the lower panel of Table \ref{tab:app_gua_results}. Using the tree-structured model results in three clusters of communities that differ in terms of their probability to use modern prenatal care. The finite mixture identifies only two   clusters. We prefer to use model selection by BIC as it showed more stable estimates in the simulations with binary response. The detected partitions and the high variance obtained by the mixed model indicate that heterogeneity of communities is definitely present. Nevertheless, only a few clusters of communities have to be distinguished. There is a strong similarity between the third cluster of the tree-structured model ($\beta_{i0}^{(3)}=1.448$) and the second cluster of the finite mixture model ($\beta_{i0}^{(2)}=1.465$) but as a whole the partition of tree-structured clustering seems to be more adequate.
In Figure \ref{fig:gua_comparison} the estimated distribution of the community-specific intercepts of the tree-structured model and the estimated normal distribution of the mixed model are graphically illustrated.

%\section{Extension: Group-specific Slopes} \label{sec:extension}

\section{Concluding Remarks}
For simplicity we focussed on the most important case of clustered intercepts. 
%So far we limited our considerations on the case of a group-specific intercept, where $z_{ij}=1$. 
However, the general fixed effects model \eqref{eq:fixedeffects_full} allows for more than one parameter to be unit-specific. It is straightforward to extend the tree-structured model to include a covariate vector $\zb_{ij}=(1,z_{ij1},\hdots,z_{ijq})$. Then one obtains a model with predictor
\begin{equation}
\eta_{ij}=\xb_{ij}^\top\betab+\sum_{r=0}^{q}\sum_{k=1}^{m_r}z_{ijr}\beta_{ir}^{(k)}I(i \in S_{kr}),
\end{equation}
where $S_{1r},\hdots,S_{m_rr}$ is a partition of the units $\{1,\hdots,n\}$ with respect to the $r$-th component of $\zb_{ij}$ and $\beta_{ir}^{(1)},\hdots,\beta_{ir}^{(m_r)}$ are the corresponding parameters of each cluster. Due to individual splits, the number and form of clusters do not have to be the same for the different components of $\zb_{ij}$. The fitting procedure given in Section \ref{sec:fitting procedure} can easily be adapted to this general model. In each iteration one simply has to determine the best split among all covariates and all corresponding splits simultaneously. In a first step the order of the units $\{1,\hdots,n\}$ with respect to single covariates has to be defined. It is not assumed that the order is the same for each of the covariates. The result is one tree for each covariate that represents a partition of units.

The proposed tree structured clustering competes well with the competitors. In particular, it performs better than the finite mixture approach and has the advantage that the number of units is not restricted as in the penalty approach. The applications were chosen to illustrate the potential of the method to find clusters that share the same effect on the covariates. The potential of the method to yield better estimates when the heterogeneity and explanatory variables are correlated is demonstrated in the simulations. The presented results were obtained by an R program that is available from the authors.

\bibliography{literatur}

\newpage

% Appendix

\section*{Appendix: Tabular Display of Simulation Results}

In the following we give the results of all settings of the simulations described in Section \ref{sec:simulations}. Each table contains the MSEs of the unit-specific intercepts, the MSEs of the linear term and the selected number of clusters as the average of 100 replications, respectively.

\newpage

% no no 0
\begin{table}[!ht]
\caption{Average results for the settings with normal response, normal distributed intercepts and $\rho=0$.}
\begin{center}
\begin{tiny}
\begin{tabular}{llrrrrrr}
\toprule
&&\multicolumn{2}{c}{\bf{MSE - intercepts}}&\multicolumn{2}{c}{\bf{MSE - linear term}}&\multicolumn{2}{c}{\bf{Number of Clusters}}\\
&&$m_0=5$&$m_0=10$&$m_0=5$&$m_0=10$&$m_0=5$&$m_0=10$ \\
\midrule
$n=200$&GFM& 2.26 & 2.26 & 0.04 & 0.04 & 200.00 & 200.00 \\
$n_i=4$&GMM& 0.68 & 0.71 & 0.03 & 0.03 & 200.00 & 200.00 \\
  &TSC& 1.56 & 1.57 & 0.04 & 0.04 & 4.96 & 5.02 \\
  &PEL&  &  &  &  &  &  \\
  &FINA& 1.05 & 1.10 & 0.03 & 0.03 & 1.89 & 1.91 \\
  &FINB& 0.99 & 1.06 & 0.03 & 0.03 & 1.31 & 1.36 \\
\midrule
$n=100$&GFM& 1.14 & 1.14 & 0.03 & 0.03 & 100.00 & 100.00 \\
$n_i=8$&GMM& 0.54 & 0.56 & 0.03 & 0.03 & 100.00 & 100.00 \\
  &TSC& 0.97 & 0.99 & 0.03 & 0.03 & 5.28 & 5.38 \\
  &PEL&  &  &  &  &  &  \\
  &FINA& 0.82 & 0.87 & 0.03 & 0.03 & 2.04 & 2.10 \\
  &FINB& 0.86 & 0.91 & 0.03 & 0.03 & 1.67 & 1.72 \\
\midrule
$n=40$&GFM& 0.45 & 0.45 & 0.03 & 0.03 & 40.00 & 40.00 \\
$n_i=20$&GMM& 0.31 & 0.32 & 0.03 & 0.03 & 40.00 & 40.00 \\
  &TSC& 0.44 & 0.46 & 0.03 & 0.03 & 5.82 & 6.00 \\
  &PEL& 0.37 & 0.38 & 0.03 & 0.03 & 15.00 & 15.06 \\
  &FINA& 0.53 & 0.55 & 0.03 & 0.03 & 2.27 & 2.44 \\
  &FINB& 0.57 & 0.61 & 0.03 & 0.03 & 1.86 & 1.98 \\
\midrule
$n=20$&GFM& 0.22 & 0.22 & 0.03 & 0.03 & 20.00 & 20.00 \\
$n_i=40$&GMM& 0.19 & 0.19 & 0.03 & 0.03 & 20.00 & 20.00 \\
  &TSC& 0.23 & 0.24 & 0.03 & 0.03 & 5.76 & 6.00 \\
  &PEL& 0.21 & 0.21 & 0.03 & 0.03 & 9.95 & 9.99 \\
  &FINA& 0.32 & 0.34 & 0.03 & 0.03 & 2.45 & 2.66 \\
  &FINB& 0.39 & 0.43 & 0.03 & 0.03 & 1.96 & 2.06 \\
\bottomrule
\end{tabular}
\end{tiny}
\end{center}
\end{table}

% no no 0.8
\begin{table}[!ht]
\caption{Average results for the settings with normal response, normal distributed intercepts and $\rho=0.8$.}
\begin{center}
\begin{tiny}
\begin{tabular}{llrrrrrr}
\toprule
&&\multicolumn{2}{c}{\bf{MSE - intercepts}}&\multicolumn{2}{c}{\bf{MSE - linear term}}&\multicolumn{2}{c}{\bf{Number of Clusters}}\\
&&$m_0=5$&$m_0=10$&$m_0=5$&$m_0=10$&$m_0=5$&$m_0=10$ \\
\midrule
$n=200$&GFM& 2.28 & 2.28 & 0.05 & 0.05 & 200.00 & 200.00 \\
$n_i=4$&GMM& 0.88 & 0.95 & 0.29 & 0.32 & 200.00 & 200.00 \\
  &TSC& 1.51 & 1.53 & 0.08 & 0.08 & 4.86 & 4.95 \\
  &PEL&  &  &  &  &  &  \\
  &FINA& 0.95 & 1.01 & 0.30 & 0.34 & 1.14 & 1.10 \\
  &FINB& 0.92 & 0.98 & 0.30 & 0.34 & 1.00 & 1.00 \\
	\midrule
$n=100$&GFM& 1.16 & 1.16 & 0.04 & 0.04 & 100.00 & 100.00 \\
$n_i=8$&GMM& 0.84 & 0.91 & 0.25 & 0.29 & 100.00 & 100.00 \\
  &TSC& 0.96 & 0.98 & 0.05 & 0.06 & 5.18 & 5.20 \\
  &PEL&  &  &  &  &  &  \\
  &FINA& 0.94 & 1.00 & 0.26 & 0.30 & 1.25 & 1.25 \\
  &FINB& 0.92 & 0.99 & 0.28 & 0.31 & 1.00 & 1.02 \\
		\midrule
$n=40$&GFM& 0.48 & 0.48 & 0.04 & 0.04 & 40.00 & 40.00 \\
$n_i=20$&GMM& 0.67 & 0.76 & 0.19 & 0.23 & 40.00 & 40.00 \\
  &TSC& 0.48 & 0.50 & 0.04 & 0.04 & 5.82 & 5.93 \\
  &PEL& 0.39 & 0.40 & 0.05 & 0.06 & 14.17 & 14.14 \\
  &FINA& 0.82 & 0.89 & 0.21 & 0.25 & 1.53 & 1.51 \\
  &FINB& 0.90 & 0.99 & 0.26 & 0.31 & 1.11 & 1.02 \\
		\midrule
$n=20$&GFM& 0.25 & 0.25 & 0.04 & 0.04 & 20.00 & 20.00 \\
$n_i=40$&GMM& 0.46 & 0.54 & 0.14 & 0.17 & 20.00 & 20.00 \\
  &TSC& 0.27 & 0.29 & 0.05 & 0.05 & 5.74 & 5.97 \\
  &PEL& 0.25 & 0.26 & 0.06 & 0.06 & 9.59 & 9.62 \\
  &FINA& 0.62 & 0.71 & 0.17 & 0.21 & 1.80 & 1.73 \\
  &FINB& 0.81 & 0.91 & 0.25 & 0.29 & 1.22 & 1.16 \\
\bottomrule
\end{tabular}
\end{tiny}
\end{center}
\end{table}

% no chi2 0
\begin{table}[!ht]
\caption{Average results for the settings with normal response, chi-squared distributed intercepts and $\rho=0$.}
\begin{center}
\begin{tiny}
\begin{tabular}{llrrrrrr}
\toprule
&&\multicolumn{2}{c}{\bf{MSE - intercepts}}&\multicolumn{2}{c}{\bf{MSE - linear term}}&\multicolumn{2}{c}{\bf{Number of Clusters}}\\
&&$m_0=5$&$m_0=10$&$m_0=5$&$m_0=10$&$m_0=5$&$m_0=10$ \\
\midrule
$n=200$&GFM& 2.27 & 2.27 & 0.04 & 0.04 & 200.00 & 200.00 \\
$n_i=4$&GMM& 0.50 & 0.59 & 0.03 & 0.03 & 200.00 & 200.00 \\
  &TSC& 1.52 & 1.59 & 0.04 & 0.04 & 4.60 & 4.88 \\
  &PEL&  &  &  &  &  &  \\
  &FINA& 0.69 & 0.77 & 0.03 & 0.03 & 1.49 & 1.80 \\
  &FINB& 0.63 & 0.76 & 0.03 & 0.03 & 1.14 & 1.32 \\
			\midrule
$n=100$&GFM& 1.10 & 1.10 & 0.03 & 0.03 & 100.00 & 100.00 \\
$n_i=8$&GMM& 0.41 & 0.47 & 0.02 & 0.02 & 100.00 & 100.00 \\
  &TSC& 0.91 & 0.95 & 0.02 & 0.03 & 4.77 & 5.14 \\
  &PEL&  &  &  &  &  &  \\
  &FINA& 0.54 & 0.50 & 0.02 & 0.02 & 1.72 & 1.90 \\
  &FINB& 0.55 & 0.55 & 0.02 & 0.02 & 1.28 & 1.53 \\
			\midrule
$n=40$&GFM& 0.45 & 0.45 & 0.03 & 0.03 & 40.00 & 40.00 \\
$n_i=20$&GMM& 0.26 & 0.28 & 0.03 & 0.03 & 40.00 & 40.00 \\
  &TSC& 0.42 & 0.42 & 0.03 & 0.03 & 4.95 & 5.15 \\
  &PEL& 0.30 & 0.29 & 0.03 & 0.03 & 13.17 & 13.27 \\
  &FINA& 0.26 & 0.28 & 0.03 & 0.03 & 1.85 & 2.00 \\
  &FINB& 0.28 & 0.29 & 0.03 & 0.03 & 1.60 & 1.68 \\
			\midrule
$n=20$&GFM& 0.23 & 0.23 & 0.03 & 0.03 & 20.00 & 20.00 \\
$n_i=40$&GMM& 0.16 & 0.16 & 0.03 & 0.03 & 20.00 & 20.00 \\
  &TSC& 0.22 & 0.23 & 0.03 & 0.03 & 4.69 & 4.92 \\
  &PEL& 0.15 & 0.15 & 0.03 & 0.03 & 7.87 & 8.23 \\
  &FINA& 0.14 & 0.18 & 0.03 & 0.03 & 1.88 & 2.10 \\
  &FINB& 0.15 & 0.20 & 0.03 & 0.03 & 1.67 & 1.81 \\
\bottomrule
\end{tabular}
\end{tiny}
\end{center}
\end{table}

% no chi2 0.8
\begin{table}[!ht]
\caption{Average results for the settings with normal response, chi-squared distributed intercepts and $\rho=0.8$.}
\begin{center}
\begin{tiny}
\begin{tabular}{llrrrrrr}
\toprule
&&\multicolumn{2}{c}{\bf{MSE - intercepts}}&\multicolumn{2}{c}{\bf{MSE - linear term}}&\multicolumn{2}{c}{\bf{Number of Clusters}}\\
&&$m_0=5$&$m_0=10$&$m_0=5$&$m_0=10$&$m_0=5$&$m_0=10$ \\
\midrule
$n=200$&GFM& 2.30 & 2.30 & 0.05 & 0.05 & 200.00 & 200.00 \\
$n_i=4$&GMM& 0.56 & 0.73 & 0.13 & 0.20 & 200.00 & 200.00 \\
  &TSC& 1.51 & 1.55 & 0.05 & 0.06 & 4.62 & 4.85 \\
  &PEL&  &  &  &  &  &  \\
  &FINA& 0.64 & 0.82 & 0.13 & 0.20 & 1.18 & 1.24 \\
  &FINB& 0.60 & 0.77 & 0.14 & 0.21 & 1.01 & 1.01 \\
				\midrule
$n=100$&GFM& 1.12 & 1.12 & 0.04 & 0.04 & 100.00 & 100.00 \\
$n_i=8$&GMM& 0.53 & 0.70 & 0.12 & 0.18 & 100.00 & 100.00 \\
  &TSC& 0.92 & 0.95 & 0.04 & 0.05 & 4.72 & 4.99 \\
  &PEL&  &  &  &  &  &  \\
  &FINA& 0.61 & 0.74 & 0.12 & 0.19 & 1.32 & 1.33 \\
  &FINB& 0.60 & 0.77 & 0.13 & 0.20 & 1.01 & 1.03 \\
				\midrule
$n=40$&GFM& 0.48 & 0.48 & 0.04 & 0.04 & 40.00 & 40.00 \\
$n_i=20$&GMM& 0.44 & 0.62 & 0.11 & 0.17 & 40.00 & 40.00 \\
  &TSC& 0.45 & 0.46 & 0.05 & 0.05 & 4.82 & 5.12 \\
  &PEL& 0.33 & 0.32 & 0.05 & 0.05 & 12.85 & 13.07 \\
  &FINA& 0.45 & 0.56 & 0.11 & 0.15 & 1.62 & 1.56 \\
  &FINB& 0.51 & 0.70 & 0.13 & 0.20 & 1.26 & 1.20 \\
				\midrule
$n=20$&GFM& 0.26 & 0.26 & 0.04 & 0.04 & 20.00 & 20.00 \\
$n_i=40$&GMM& 0.30 & 0.44 & 0.08 & 0.13 & 20.00 & 20.00 \\
  &TSC& 0.26 & 0.26 & 0.04 & 0.04 & 4.69 & 4.92 \\
  &PEL& 0.20 & 0.19 & 0.04 & 0.04 & 8.04 & 8.21 \\
  &FINA& 0.31 & 0.44 & 0.08 & 0.11 & 1.74 & 1.77 \\
  &FINB& 0.38 & 0.62 & 0.11 & 0.17 & 1.39 & 1.34 \\
\bottomrule
\end{tabular}
\end{tiny}
\end{center}
\end{table}

% bin no 0
\begin{table}[!ht]
\caption{Average results for the settings with binary response, normal distributed intercepts and $\rho=0$.}
\begin{center}
\begin{tiny}
\begin{tabular}{llrrrrrr}
\toprule
&&\multicolumn{2}{c}{\bf{MSE - intercepts}}&\multicolumn{2}{c}{\bf{MSE - linear term}}&\multicolumn{2}{c}{\bf{Number of Clusters}}\\
&&$m_0=5$&$m_0=10$&$m_0=5$&$m_0=10$&$m_0=5$&$m_0=10$ \\
\midrule
$n=100$&GFM&   &  &  &  &  &   \\
$n_i=8$&GMM& 0.74 & 0.88 & 0.03 & 0.03 & 100.00 & 100.00 \\
  &TSC& 1.06 & 1.29 & 0.02 & 0.02 & 2.96 & 2.98 \\
  &PEL&  &  &  &  &  &  \\
  &FINA& 2.88 & 2.39 & 0.03 & 0.03 & 2.98 & 3.03 \\
  &FINB& 2.11 & 1.66 & 0.03 & 0.02 & 2.64 & 2.63 \\
					\midrule
$n=40$&GFM&   &  &  &  &  &   \\
$n_i=20$&GMM& 0.48 & 0.56 & 0.02 & 0.02 & 40.00 & 40.00 \\
  &TSC& 0.70 & 0.87 & 0.02 & 0.02 & 3.32 & 3.50 \\
  &PEL& 1.23 & 1.20 & 0.02 & 0.02 & 10.78 & 14.28 \\
  &FINA& 10.70 & 5.26 & 0.02 & 0.02 & 3.49 & 3.52 \\
  &FINB& 9.10 & 3.93 & 0.02 & 0.02 & 3.00 & 2.97 \\
					\midrule
$n=20$&GFM&   &  &  &  &  &   \\
$n_i=40$&GMM& 0.71 & 0.62 & 0.03 & 0.03 & 20.00 & 20.00 \\
  &TSC& 2.40 & 2.18 & 0.03 & 0.03 & 3.44 & 3.84 \\
  &PEL& 1.44 & 1.15 & 0.03 & 0.03 & 5.70 & 9.15 \\
  &FINA& 19.94 & 12.58 & 0.03 & 0.03 & 3.57 & 3.84 \\
  &FINB& 15.58 & 8.71 & 0.03 & 0.03 & 3.12 & 3.21 \\
\bottomrule
\end{tabular}
\end{tiny}
\end{center}
\end{table}

% bin no 0.8
\begin{table}[!ht]
\caption{Average results for the settings with binary response, normal distributed intercepts and $\rho=0.8$.}
\begin{center}
\begin{tiny}
\begin{tabular}{llrrrrrr}
\toprule
&&\multicolumn{2}{c}{\bf{MSE - intercepts}}&\multicolumn{2}{c}{\bf{MSE - linear term}}&\multicolumn{2}{c}{\bf{Number of Clusters}}\\
&&$m_0=5$&$m_0=10$&$m_0=5$&$m_0=10$&$m_0=5$&$m_0=10$ \\
\midrule
$n=100$&GFM&   &  &  &  &  &   \\
$n_i=8$&GMM& 2.13 & 2.55 & 0.48 & 0.54 & 100.00 & 100.00 \\
  &TSC& 1.59 & 1.93 & 0.25 & 0.29 & 2.46 & 2.38 \\
  &PEL&  &  &  &  &  &  \\
  &FINA& 3.43 & 3.89 & 0.46 & 0.51 & 2.35 & 2.26 \\
  &FINB& 2.60 & 2.95 & 0.50 & 0.56 & 1.93 & 1.85 \\
						\midrule
$n=40$&GFM&   &  &  &  &  &   \\
$n_i=20$&GMM& 0.92 & 1.12 & 0.14 & 0.15 & 40.00 & 40.00 \\
  &TSC& 0.98 & 1.16 & 0.11 & 0.12 & 3.04 & 3.13 \\
  &PEL& 1.32 & 1.26 & 0.05 & 0.05 & 10.42 & 13.19 \\
  &FINA& 12.51 & 8.08 & 0.11 & 0.14 & 2.96 & 2.91 \\
  &FINB& 8.06 & 5.39 & 0.16 & 0.22 & 2.45 & 2.29 \\
						\midrule
$n=20$&GFM&   &  &  &  &  &   \\
$n_i=40$&GMM& 0.87 & 0.84 & 0.07 & 0.08 & 20.00 & 20.00 \\
  &TSC& 2.67 & 1.87 & 0.06 & 0.07 & 3.21 & 3.53 \\
  &PEL& 1.74 & 1.26 & 0.05 & 0.05 & 5.61 & 8.91 \\
  &FINA& 22.57 & 13.19 & 0.06 & 0.09 & 3.34 & 3.41 \\
  &FINB& 15.15 & 7.81 & 0.09 & 0.14 & 2.81 & 2.64 \\
\bottomrule
\end{tabular}
\end{tiny}
\end{center}
\end{table}

% bin chi2 0
\begin{table}[!ht]
\caption{Average results for the settings with binary response, chi-squared distributed intercepts and $\rho=0$.}
\begin{center}
\begin{tiny}
\begin{tabular}{llrrrrrr}
\toprule
&&\multicolumn{2}{c}{\bf{MSE - intercepts}}&\multicolumn{2}{c}{\bf{MSE - linear term}}&\multicolumn{2}{c}{\bf{Number of Clusters}}\\
&&$m_0=5$&$m_0=10$&$m_0=5$&$m_0=10$&$m_0=5$&$m_0=10$ \\
\midrule
$n=100$&GFM&   &  &  &  &  &   \\
$n_i=8$&GMM& 0.68 & 0.92 & 0.02 & 0.02 & 100.00 & 100.00 \\
  &TSC& 0.91 & 1.39 & 0.02 & 0.02 & 2.79 & 2.85 \\
  &PEL&  &  &  &  &  &  \\
  &FINA& 1.72 & 2.30 & 0.02 & 0.02 & 2.74 & 2.90 \\
  &FINB& 1.42 & 1.70 & 0.02 & 0.02 & 2.40 & 2.51 \\
							\midrule
$n=40$&GFM&   &  &  &  &  &   \\
$n_i=20$&GMM& 0.48 & 0.61 & 0.02 & 0.02 & 40.00 & 40.00 \\
  &TSC& 0.59 & 0.82 & 0.02 & 0.02 & 3.01 & 3.37 \\
  &PEL& 1.60 & 1.43 & 0.02 & 0.02 & 9.83 & 12.35 \\
  &FINA& 5.61 & 6.94 & 0.02 & 0.02 & 3.04 & 3.34 \\
  &FINB& 4.66 & 4.47 & 0.02 & 0.02 & 2.74 & 2.91 \\
							\midrule
$n=20$&GFM&   &  &  &  &  &   \\
$n_i=40$&GMM& 1.61 & 2.00 & 0.03 & 0.03 & 20.00 & 20.00 \\
  &TSC& 2.81 & 2.96 & 0.03 & 0.03 & 2.94 & 3.56 \\
  &PEL& 1.93 & 2.04 & 0.03 & 0.02 & 5.75 & 8.04 \\
  &FINA& 21.18 & 19.61 & 0.04 & 0.03 & 3.04 & 3.61 \\
  &FINB& 19.95 & 16.30 & 0.03 & 0.03 & 2.77 & 3.06 \\
\bottomrule
\end{tabular}
\end{tiny}
\end{center}
\end{table}

% bin chi2 0.8
\begin{table}[!ht]
\caption{Average results for the settings with binary response, chi-squared distributed intercepts and $\rho=0.8$.}
\begin{center}
\begin{tiny}
\begin{tabular}{llrrrrrr}
\toprule
&&\multicolumn{2}{c}{\bf{MSE - intercepts}}&\multicolumn{2}{c}{\bf{MSE - linear term}}&\multicolumn{2}{c}{\bf{Number of Clusters}}\\
&&$m_0=5$&$m_0=10$&$m_0=5$&$m_0=10$&$m_0=5$&$m_0=10$ \\
\midrule
$n=100$&GFM&   &  &  &  &  &   \\
$n_i=8$&GMM& 1.55 & 2.30 & 0.41 & 0.50 & 100.00 & 100.00 \\
  &TSC& 1.28 & 1.91 & 0.22 & 0.30 & 2.50 & 2.30 \\
  &PEL&  &  &  &  &  &  \\
  &FINA& 4.85 & 4.37 & 0.33 & 0.46 & 2.48 & 2.24 \\
  &FINB& 2.68 & 2.45 & 0.37 & 0.50 & 2.05 & 1.86 \\
								\midrule
$n=40$&GFM&   &  &  &  &  &   \\
$n_i=20$&GMM& 0.72 & 1.15 & 0.13 & 0.16 & 40.00 & 40.00 \\
  &TSC& 0.75 & 1.15 & 0.09 & 0.12 & 2.80 & 3.01 \\
  &PEL& 1.72 & 1.53 & 0.04 & 0.05 & 9.38 & 11.89 \\
  &FINA& 9.57 & 6.76 & 0.09 & 0.14 & 2.85 & 2.84 \\
  &FINB& 7.06 & 4.68 & 0.11 & 0.17 & 2.50 & 2.47 \\
								\midrule
$n=20$&GFM&   &  &  &  &  &   \\
$n_i=40$&GMM& 1.66 & 2.26 & 0.07 & 0.07 & 20.00 & 20.00 \\
  &TSC& 3.08 & 2.92 & 0.06 & 0.07 & 2.81 & 3.33 \\
  &PEL& 2.26 & 2.34 & 0.05 & 0.05 & 5.59 & 7.72 \\
  &FINA& 21.87 & 21.18 & 0.06 & 0.08 & 2.90 & 3.25 \\
  &FINB& 21.79 & 16.13 & 0.07 & 0.11 & 2.68 & 2.68 \\
\bottomrule
\end{tabular}
\end{tiny}
\end{center}
\end{table}

\end{document}